\documentclass[%
superscriptaddress,
aps,
prl, 
twocolumn,
showpacs,
longbibliography,
notitlepage,
floatfix,
]{revtex4-2}

\usepackage{mathtools}
\usepackage{amsbsy}
\usepackage{amstext}
\usepackage{amsmath}
\usepackage{amssymb}
\usepackage{txfonts}
\usepackage{color}
\usepackage{graphicx}
\usepackage{wrapfig}
\usepackage{float}
\usepackage[normalem]{ulem}
\usepackage{physics}
\usepackage{bbold}
\usepackage{cancel}
\usepackage{bm}

\usepackage[version=4]{mhchem} 
\usepackage{xcolor} 
\usepackage{multirow} 
\usepackage[caption=false]{subfig}

\usepackage{tikz}
\usetikzlibrary{quantikz}

\usepackage[unicode=true,pdfusetitle,
 bookmarks=false,
 breaklinks=false,
 pdfborder={0 0 1},
 backref=false,
 colorlinks=true, linkcolor=magenta, urlcolor=gray, citecolor=olive, filecolor=blue]
 {hyperref}

\hypersetup{
 bookmarksnumbered=false,
 bookmarksopen=false}

\makeatletter

\@ifundefined{textcolor}{}{
 \definecolor{BLACK}{gray}{0}
 \definecolor{WHITE}{gray}{1}
 \definecolor{RED}{rgb}{1,0,0}
 \definecolor{GREEN}{rgb}{0,0.7,0}
 \definecolor{BLUE}{rgb}{0,0,1}
 \definecolor{CYAN}{cmyk}{1,0,0,0}
 \definecolor{MAGENTA}{cmyk}{0,1,0,0}
 \definecolor{YELLOW}{cmyk}{0,0,1,0}
}

\newcommand{\beq}{\begin{equation}}
\newcommand{\eeq}{\end{equation}}
\newcommand{\beqa}{\begin{eqnarray}}
\newcommand{\eeqa}{\end{eqnarray}}

\let\origaddcontentsline\addcontentsline
\newcommand{\suppressTOC}{\let\addcontentsline\@gobblethree}
\newcommand{\restoreTOC}{\let\addcontentsline\origaddcontentsline}

\makeatother
\begin{document}

\preprint{APS/123-QED}

\suppressTOC

\title{Intrinsic Vectorial Gradiometry via Quantum Control of a Spin-based Sensor}

\author{Jaime García Oliván}\email{Contact author: jaime.garcia@ehu.eus}
\affiliation{Department of Physical Chemistry, University of the Basque Country UPV/EHU, Apartado 644, 48080 Bilbao, Spain}
\affiliation{EHU Quantum Center, University of the Basque Country UPV/EHU, 48080 Bilbao, Spain}

\author{Pablo Acedo}
\affiliation{EHU Quantum Center, University of the Basque Country UPV/EHU, 48080 Bilbao, Spain}
\affiliation{Department of Physics, University of the Basque Country UPV/EHU, Apartado 644, 48080 Bilbao, Spain}
\affiliation{IKERBASQUE, Basque Foundation for Science, 48011 Bilbao, Spain}

\author{Oliver T. Whaites}
\affiliation{Department of Physical Chemistry, University of the Basque Country UPV/EHU, Apartado 644, 48080 Bilbao, Spain}
\affiliation{EHU Quantum Center, University of the Basque Country UPV/EHU, 48080 Bilbao, Spain}

\author{Jorge Casanova}
\affiliation{Department of Physical Chemistry, University of the Basque Country UPV/EHU, Apartado 644, 48080 Bilbao, Spain}
\affiliation{EHU Quantum Center, University of the Basque Country UPV/EHU, 48080 Bilbao, Spain}

\date{\today}

\begin{abstract}
    Gradiometry provides a versatile alternative to passive environmental shielding in quasi-static magnetometry, effectively suppressing background noise through differential signal extraction. Nevertheless, traditional implementations rely on multi-sensor architectures restricted to spatial gradients, where subtracting signals from independent detectors involves imperfect suppression of common-mode noise and artifacts, limiting their sensitivity. To overcome these limitations, we introduce a quantum control sequence that enables intrinsic temporal and spatial vectorial gradiometry of magnetic fields using a single quantum sensor. Our method provides direct access to first  and higher-order  derivatives of the magnetic field and extended applicability via auxiliary nuclear spin memory. We showcase this protocol on an ensemble of nitrogen-vacancy (NV) centers in diamond and combine it with mechanical control to realize high-precision differential sensing. Through detailed numerical simulations, we demonstrate the performance of our scheme in two critical DC magnetometry applications: (i) vector magnetic anomaly detection and (ii) non-invasive gradiometry of neuronal action potentials.  
\end{abstract}

\maketitle

\section{Introduction}

Highly sensitive magnetometers enable a wide range of applications across diverse fields~\cite{Alsina_2025, Grafenstein_2025, Katsumi_2025, Whaites_2026}. In particular, quasi-static and low-frequency field sensing facilitates  geological surveying~\cite{Dang_2010,Gang_2016,Newman_2024,Levi_2025}, nanoscopic characterization of magnetic materials~\cite{Huxter_2022,Huxter_2023,Melendez_2025} and measurement of biological signals~\cite{Barry_2016,Webb_2021,Arai_2022,Yu_2024,Omar_2026}. Standard DC magnetometry techniques rely on direct field intensity measurements, making them highly susceptible to magnetic drifts over extended measurement period \cite{Huxter_2022, Wood_2022, Richter_2026} as well as to strong static background fields, such as the geomagnetic field \cite{Zhang_2023,Alem_2023}. These heavily disturb the sensor, degrading its sensitivity, such that shielding from the environment is often introduced to suppress them. However, in certain relevant applications, \textit{e.g.}, detection of magnetic anomalies, the system cannot be isolated from its surroundings. 

Gradiometers offer an elegant solution to these issues. By measuring a differential signal, static background fields and magnetic drifts are suppressed. Gradiometry has been demonstrated across a wide range of quantum platforms, including superconducting quantum interference devices (SQUIDs)~\cite{Dang_2010,Chwala_2019,Gkika_2024} and optical pumping magnetometers (OPMs) \cite{Sheng_2017,Cook_2024}. Typically, these reconstruct the magnetic field gradient  by combining the readout from two (or more) spatially separated, independent sensors. However, this common-mode effect rejection is often imperfect, resulting in added noise which is detrimental for the sensitivity of the gradiometer. Furthermore, multi-sensor gradiometers show a worsening in the overall sensitivity compared to single-sensor measurements \cite{Zhang_2016, Zhang_2020}.

Intrinsic gradiometers, on the other hand, provide a more robust alternative. Previous studies based on OPMs have demonstrated intrinsic spatial gradiometry using multi-cell and single-cell atomic ensembles by measuring light polarization after interacting with two independent cells or two regions of a single, large cell, respectively~\cite{Zhang_2020,Zilinski_2026}. In the former case, localized environmental fluctuations can degrade performance across sensors, whereas the latter avoids this issue but requires a larger cell volume.

\begin{figure}[b!]
    \centering
    \includegraphics[width=0.48\textwidth]{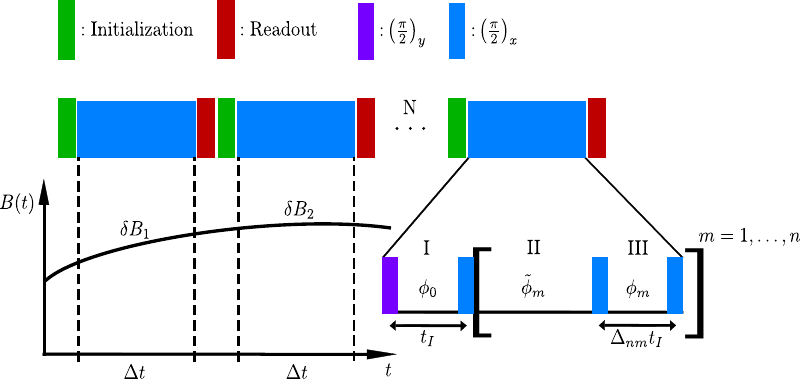}
    \caption{Gradiometric sensing scheme. Our protocol comprises an initialization and a readout step, and a measurement sequence (in green, red and blue, respectively), sequentially repeated $N$ times. Our sequence allows to measure the magnetic field difference between two points separated by a time $\Delta t$. Considering that the magnetic field $B(t)$ changes slowly within a single sequence, each measurement will be proportional to the field difference within the sequence, $\delta B_i = B(t_{i} + \Delta t) - B(t_{i})$, such that, after $N$ repetitions, the full gradient of the field can be recovered. Here, $t_i$ sets the start of $i$-th measurement. Each of these sequences consists of two interrogation stages (I and III) wherein the value of the field remains unchanged and a signal evolution stage (II) wherein the field changes its value. Repetition of stages II and III within each sequence allows access to $n$-th order derivative of the field. Here, the duration of stage III in the $m$-th block (relative to stage I) is given by the combinatorial number $\Delta_{nm} = \left(\begin{array}{c}
    n \\
    m \\
\end{array}\right)$.}
    \label{F: Gradiometry}
\end{figure}

In the case of solid-state spin defects, the standard approach for DC magnetometry is Ramsey interferometry~\cite{Taylor_2008}. In the particular case of color centers, DC magnetometry via optically detected magnetic resonance (ODMR) has recently gained interest \cite{Dreau_2011,Schloss_2018,Voce_2025}. Such measurements are particularly vulnerable to environmental noise when probing weak signals in complex environments, to the extent that high-precision applications --such as magnetoneurography and magnetomyography-- typically require heavy magnetic shielding~\cite{Zhang_2023}. 

Spatial gradiometers using color centers have been implemented using pairs of non-interacting sensors~\cite{Blakley_2018,Masuyama_2021,Zhang_2023,Omar_2026}, though these configurations are prone to  errors arising from imperfect suppression of common-mode artifacts. Conversely, recent intrinsic spatial gradiometry with a single NV center mounted on an oscillating cantilever tip~\cite{Huxter_2022,Huxter_2023} bypasses multi-sensor noise, but relies on extracting differential readouts  at the oscillation extrema. This rigidly locks the spatial gradient baseline to the physical amplitude of the tip motion, which  confines the technique to the nanoscale, rendering it challenging to adapt to broader spatial baselines. Furthermore, to the best of our knowledge, no intrinsic gradiometer has been proposed for time domain measurements.

In this work, we propose a quantum control scheme which enables intrinsic temporal and spatial vectorial gradiometry with a quantum sensor (see Fig. \ref{F: Gradiometry}) while suppressing background fields and magnetic drifts. Notably, our protocol provides access to first, and higher-order derivatives of the magnetic field and can naturally incorporate a nuclear memory to achieve extended operational times. The manuscript is structured as follows. In Section \ref{S: Protocol} we present a detailed description of our protocol. Next, in Section \ref{S: DCtoAC}, we investigate a method to convert DC magnetic fields into effective AC signal via physical motion of the sensor in order to improve the sensitivity of our protocol. We conduct numerical simulations to showcase our sensing scheme and present the results in Section \ref{S: Results}. In particular, our simulations model a diamond sensor containing an ensemble of NV centers, which intrinsically incorporate a nuclear spin memory. Leveraging this, we investigate two important application for DC magnetometry: vectorial magnetic anomaly detection and measurement of magnetic field gradients originating from neuronal action potentials (APs). Our results demonstrate the broad applicability of the protocol for highly sensitive gradiometry. Finally, in Section \ref{S: Discussion} we discuss and conclude.

\section{Pulse Sequence for Intrinsic Gradiometry}\label{S: Protocol}

Our sequence comprises two interrogation stages (I and III) separated by a signal evolution stage (II) (see Fig.~\ref{F: Gradiometry}). Here, all pulses are applied along the same axis (\textit{e.g.}, $x$) except the first, which is applied along an orthogonal direction, see blue block in Fig.~\ref{F: Gradiometry}. As long as the phase accumulated by the sensor during the intermediate stage is small, the measured signal will be proportional to the difference between the field values at both interrogation stages. By adding blocks of signal evolution and sensor interrogation stages (see Fig. \ref{F: Gradiometry}), one recovers higher order derivatives of the target field. In the following, we describe in detail the theory behind our pulse sequence. 

The theoretical description that follows is general for any pair of electron and nuclear spin qubits, which we call the sensor ($e$) and the ancilla ($a$) or memory, and are both initialized in the $|0\rangle$ state, such that $\rho_e = \rho_a = \frac{\mathbb{1} + \sigma_z}{2}$. The signal to be detected $b(t)$ is weak and slowly varying, hidden behind a static background field $B$. For the sake of simplicity in the presentation, we consider the time dependent signal $b(t)$ to remain static during stages I and III, but not during stage II, although this need not always be the case. Repetition of the sequence enables the discretization of the field, as shown in Fig. \ref{F: Gradiometry}, such that it takes values $b_I^i = b(t_i)$ and $b_{III}^i = b(t_i + \Delta t)$ during stages I and III in the $i$-th repetition, respectively, where $t_i$ sets the start of the $i$-th sequence and $\Delta t = t_{II}$ is the duration of stage II. Then, the $i$-the measurement will be proportional to the field difference in that sequence, \textit{i.e.}, $\delta b_i = b(t_i + \Delta t) - b(t_i)$, from where the discretized temporal gradient of the total field $B(t) = B + b(t)$ can be found, $\left.\frac{dB(t)}{dt}\right|_{t_i} \approx \frac{\delta b_i}{\Delta t}$. The temporal behavior described directly models relevant practical scenarios detailed in subsequent sections.

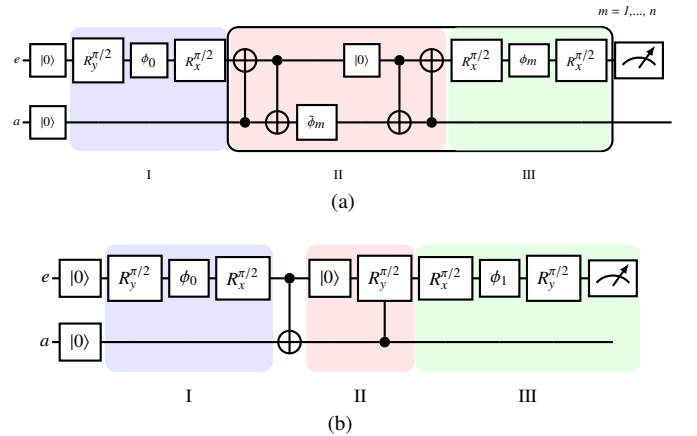
\begin{figure}[b!]
\subfloat[\label{F: 1b}]{%
  \begin{quantikz}[row sep=0.3cm, column sep=0.1cm, inner xsep=0.05pt,
    inner ysep=0.05pt, font=\tiny] 
    \lstick{\textit{e}} 
    & \gate{\ket{0}}
    & \gate{R_y^{\pi/2}}\gategroup[wires=2,steps=3,style={fill=blue!20, inner xsep=1pt, draw=none, fill opacity=0.5, rounded corners}, background, label style={label position=below,anchor=center,yshift=-0.5cm}]{\text{I}}
    & \gate{\phi_0}
    & \gate{R_x^{\pi/2}}
    & \targ{}\gategroup[wires=2,steps=6,style={fill=red!20, inner xsep=1pt, draw=none, fill opacity=0.5, rounded corners}, background, label style={label position=below,anchor=center,yshift=-0.5cm}]{\text{II}}\gategroup[wires=2,steps=9,style={inner xsep=2pt, draw=black, fill opacity=0.5, rounded corners}, background, label style={label position=above right,anchor=east,yshift=0cm,xshift=2.3cm}]{\text{\textit{m = 1,..., n}}}
    & \ctrl{1}
    & \qw
    & \gate{\ket{0}}
    & \ctrl{1}
    & \targ{}
    & \gate{R_{x}^{\pi/2}}\gategroup[wires=2,steps=3,style={fill=green!20, inner xsep=1pt, draw=none, fill opacity=0.5, rounded corners}, background, label style={label position=below,anchor=center,yshift=-0.5cm}]{\text{III}}
    & \gate{\phi_m}
    & \gate{R_{x}^{\pi/2}}
    & \meter{}
    \\
    \lstick{\textit{a}}
    & \gate{\ket{0}}
    & \qw
    & \qw
    & \qw
    & \ctrl{-1}
    & \targ{}
    & \gate{\tilde{\phi}_m}
    & \qw
    & \targ{}
    & \ctrl{-1}
    & \qw
    & \qw
    & \qw
    & \qw
    & \qw
\end{quantikz}
}\hfill%
\subfloat[\label{F: 1c}]{%
  \begin{quantikz}[row sep=0.3cm, column sep=0.1cm, inner xsep=0.05pt,
    inner ysep=0.05pt, font=\scriptsize]
    \lstick{\textit{e}} 
    & \gate{\ket{0}}
    & \gate{R_y^{\pi/2}}\gategroup[wires=2,steps=3,style={fill=blue!20, inner xsep=1pt, draw=none, fill opacity=0.5, rounded corners}, background, label style={label position=below,anchor=center,yshift=-0.5cm}]{\text{I}}
    & \gate{\phi_0}
    & \gate{R_x^{\pi/2}}
    & \ctrl{1}
    & \gate{\ket{0}}\gategroup[wires=2,steps=2,style={fill=red!20, inner xsep=1pt, draw=none, fill opacity=0.5, rounded corners}, background, label style={label position=below,anchor=center,yshift=-0.5cm}]{\text{II}}
    & \gate{R_y^{\pi/2}}
    & \gate{R_{x}^{\pi/2}}\gategroup[wires=2,steps=4,style={fill=green!20, inner xsep=1pt, draw=none, fill opacity=0.5, rounded corners}, background, label style={label position=below,anchor=center,yshift=-0.5cm}]{\text{III}}
    & \gate{\phi_1}
    & \gate{R_{y}^{\pi/2}}
    & \meter{}
    \\
    \lstick{\textit{a}}
    & \gate{\ket{0}}
    & \qw
    & \qw
    & \qw
    & \targ{}
    & \qw
    & \ctrl{-1}
    & \qw
    & \qw
    & \qw
    & \qw
\end{quantikz}}
\caption{Circuit diagram of the memory assisted pulse sequence for intrinsic temporal gradiometry. (a) General sequence for measuring the $n$-th order derivative of the field. SWAP gates may be used during stage II for encoding/recovering the sensor ($e$) state in/from a nuclear ancillary qubit ($a$) such that the protocol is limited by the nuclear dephasing time $T_2^a$. (b) Modified protocol for $T_1^a$-limited magnetometry. Here, only the population information is encoded in the nuclear memory during stage II.} \label{F: Protocol}
\end{figure}

Our protocol allows for the construction of this discretized gradient as follows: during the $i$-th measurement block in Fig.~\ref{F: Gradiometry}, the sensor first accumulates a phase $\phi_0 \propto B + b_I^i$, where $b_I^i = b(t_i)$ is the value of the signal $b(t)$ during stage I. Next, in stage II, free evolution of the sensor during signal variation results in accumulation of a phase $\tilde{\phi}_1$. Lastly, the sensor acquires a phase $\phi_1 \propto B + b_{III}^i$, where $b_{III}^i = b(t_i + \Delta t)$, is the value of the signal during stage III. Upon measurement of the spin state of the sensor, we find the expectation value of the third Pauli matrix $\sigma_z$ to be
\begin{equation}\label{Eq. Gradiometry}
    \langle \sigma_z \rangle = \cos\phi_0 \sin\phi_1 - \cos\tilde{\phi}_1 \sin\phi_0 \cos\phi_1 \approx \sin(\phi_1 - \phi_0).
\end{equation}
It follows from Eq. (\ref{Eq. Gradiometry}) that the static environmental noise $B$ is completely removed, and because the field difference between stages I and III is expected to be small, the expectation value becomes linear in the accumulated phases, such that we can write $\langle \sigma_z \rangle \approx \phi_1 - \phi_0 \propto b_{III}^i - b_I^i = \delta b_i$. As described above, by performing $N$ of such measurements, one can find the discretized temporal gradient of the total field $B(t)$. For the condition in Eq. (\ref{Eq. Gradiometry}) to be fulfilled, the duration of stage II needs to be smaller than the sensor dephasing time $T_2^*$. However, because this stage is not used for measurement, dynamical decoupling (DD) techniques can be applied such that this limitation is extended to the coherence time $T_2$. 

Interestingly, our sequence can be generalized for measuring the $n$-th order derivative. Here, the block comprising stages II and III is repeated $n$ times and we allow a variable duration of stage III in the $m$-th repetition (see Fig. \ref{F: Gradiometry}). Then, the state of the sensor at the end of the $m$-th block reads:
\begin{equation}
    \rho_e^{(m)} = \frac{\mathbb{1}}{2} + \frac{1}{2}(a_m \sigma_x + b_m \sigma_y + c_m \sigma_z),
\end{equation}
where the coefficients $a_m$, $b_m$ and $c_m$ are given recursively by
\begin{equation}
    \begin{array}{ccl}
        a_m & = & \cos\phi_m \cos\tilde{\phi}_m a_{m-1} - \cos\phi_m \sin\tilde{\phi}_m b_{m-1} + \sin\phi_m c_{m-1}, \\
        b_m & = & -\sin\tilde{\phi}_m a_{m-1} - \cos\tilde{\phi}_m b_{m-1}, \\
        c_m & = & \sin\phi_m \cos\tilde{\phi}_m a_{m-1} - \sin\phi_m \sin\tilde{\phi}_m b_{m-1} - \cos\phi_m c_{m-1},
    \end{array}
\end{equation}
with $a_0 = \cos\phi_0$, $b_0 = 0$ and $c_0 = \sin\phi_0$. Remarkably, we find that $\langle \sigma_z \rangle^{(n)} = c_n$, which assuming small $\tilde{\phi}_m$ reads (see Section \ref{SM: Gradiometry} in the Supplementary Information for more details):
\begin{equation}\label{Eq. Derivatives}
    \langle \sigma_z \rangle^{(n)} \approx (-1)^n \sin\left[\sum_{m=0}^n (-1)^m \phi_m\right].
\end{equation}
Note that Eq. (\ref{Eq. Derivatives}) is the generalization of Eq. (\ref{Eq. Gradiometry}) to derivatives of any order $n$.

To achieve the form of the $n$-th discretized derivative, the duration of the $m$-th stage III is set to $t_{III, m}^{(n)} = \Delta_{nm} t_I$, where $\Delta_{nm} = \left(\begin{array}{c}
n \\
m \\
\end{array}\right)$ is the combinatorial number. Again, we see that the background field $B$ is suppressed and then, because the total phase accumulated is expected to be small, Eq. (\ref{Eq. Derivatives}) is rewritten as $\langle \sigma_z \rangle^{(n)} \approx \sum_{m=0}^n (-1)^{n+m} \phi_m \propto (-1)^n b_I^i + \sum_{m=1}^n (-1)^{n+m} \Delta_{nm} b_{III, m}^i = \delta b^{(n)}_i$, where $b_I^i$ and $b_{III, m}^i$ are the values of the signal $b(t)$ during stage I and the $m$-th stage III, respectively.  As before, by performing $N$ of such measurements, the discretized $n$-th order time derivative of the total field $B(t)$ can be obtained: $\left.\frac{d^{(n)} B (t)}{dt^n}\right|_{t_i} \approx \frac{\delta b_i^{(n)}}{\Delta t^n}$, with the superscript $i$ referring to the $i$-th measurement. As an example, consider the second order protocol; here, we find $\langle \sigma_z \rangle^{(2)} \propto b_I^i - 2b_{III, 1}^i + b_{III, 2}^i = \delta b_i^{(2)}$, such that $\left. \frac{d^{(2)}B(t)}{dt^2}\right|_{t_i} \frac{b_I^i - 2b_{III, 1}^i + b_{III, 2}^i}{t_{II}^2}$ is the discretized second order derivative of the field $B(t)$.

As discussed above, the duration of stage II cannot be longer than the sensor coherence time for the approximations in Eqs. (\ref{Eq. Gradiometry}) and (\ref{Eq. Derivatives}) to remain valid. To surpass this limitation, stage II can be transfered to the state of the memory. By applying SWAP gates at the beginning and end of stage II, the state of the sensor can be encoded in and then recovered from the memory (see Fig. \ref{F: 1b}). An efficient construction of the SWAP gates consists in applying double CNOT gates as follows: $\mathrm{C}_a\mathrm{NOT}_e - \mathrm{C}_e\mathrm{NOT}_a$ gates to encode the state and $\mathrm{C}_e\mathrm{NOT}_a - \mathrm{C}_a\mathrm{NOT}_e$ to recover it. Such storage is benefitial because the gyromagnetic ratio of a nuclear ancilla is much smaller than that of the electron, $\gamma_a / \gamma_e \sim 10^{-3}$, such that the adverse phase accumulated by the memory during stage II will be much smaller than if it was accumulated by the sensor. Thus, the protocol is now limited by the larger nuclear dephasing time $T_2^a \sim T_1^e$, enabling longer durations of stage II and thus improves the applicability of the protocol.

A final modification to the first order gradiometry protocol can be made such that stage II is limited by the nuclear relaxation time $T_1^a$, which is at least one order of magnitude longer than the dephasing time $T_2^a$. Instead of the first SWAP gate, a $\mathrm{C}_e\mathrm{NOT}_a$ gate is applied before stage II, such that only the information about the sensor qubit populations is transfered to the nuclear memory. Thus, no phase is accumulated during this stage. The sensor is reinitialized to the $|0\rangle$ state and then a $\mathrm{C}_a\mathrm{ROT}_e(\pi/2)$ gate along the $y$-axis is applied to recover the state from the memory (see Fig. \ref{F: 1c}). Measuring the sensor state, we find
\begin{equation}
    \langle \sigma_z \rangle = -\frac{1}{2}\left\{ \cos\phi_1 - \sin\phi_0 \cos\phi_1 + \sin\phi_1 + \sin\phi_0 \sin\phi_1 \right\}.
\end{equation}
Here, as long as $\phi_0$ and $\phi_1$ are small enough as to neglect the term $\sin\phi_0 \sin\phi_1$, we may write $\langle \sigma_z \rangle \approx -1/2\left\{ 1 + (\phi_1 - \phi_0)\right\}$. 

Although this modified sequence does not allow measurement of higher order derivatives, it can be used for measuring deviations with respect to a reference signal. In this case, the state of the memory is unaffected by the CROT operation, such that the information about the first accumulated phase $\phi_0$ remains in the nuclear state. Thus, subsequent measurements after repetition of stages II and III will be proportional to the difference between $\phi_0$ and some phase $\phi_1$ accumulated due to the field at a later time. This can be use for signal referencing, \textit{e.g.}, to monitor magnetic drifts with respect to some initial magnetic field.

In all previous cases, the interrogation stages I and III, Ramsey sequences, are limited by the short dephasing time $T_2^*$ of the sensor. However, the conversion of a DC field into an eAC signal via fast mechanical control \cite{Wood_2018, Wood_2021_1, Wood_2021_2, Wood_2022}, ancillary qubits \cite{Liu_2019} or flux modulation \cite{Xie_2022}, enables the application of dynamical decoupling techniques which lead to sensor interrogation times limited by the much longer coherence time $T_2$. In this work, we investigate the DC to AC conversion via fast mechanical oscillation of the sensor by placing it on the tip of a vibrating cantilever (see Fig. \ref{F: System}).

\section{DC to AC Conversion} \label{S: DCtoAC}

\begin{figure}[t]
    \centering
    \includegraphics[width=0.45\textwidth]{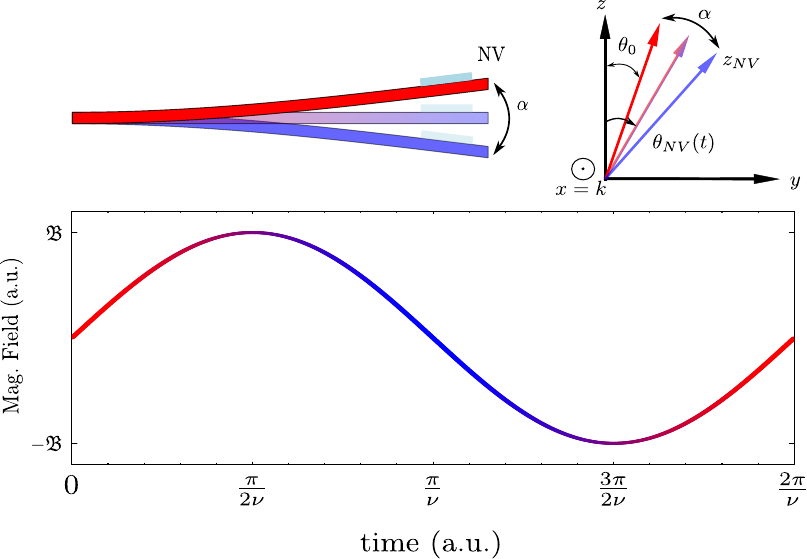}
    \caption{Schematic of DC to AC conversion method. (Top left) An NV-based sensor is placed on the tip of an vibrating cantilever, where the oscillation amplitude is $\alpha$. (Top right) The orientation of the sensor coordinate system changes with respect to the laboratory frame following the oscillatory motion of the cantilever. (Bottom) Imprinted eAC signal over a full oscillation of the cantilever. The amplitude of this signal is proportional to the intensity of the DC field, $\mathfrak{B} = \alpha \cos\theta_0 B_y$, and its frequency to the cantilever vibration frequency, $\nu$.}
    \label{F: System}
\end{figure}

\begin{figure*}[t!]
    \subfloat[\label{F:3a}]{%
    \includegraphics[width=0.32\textwidth]{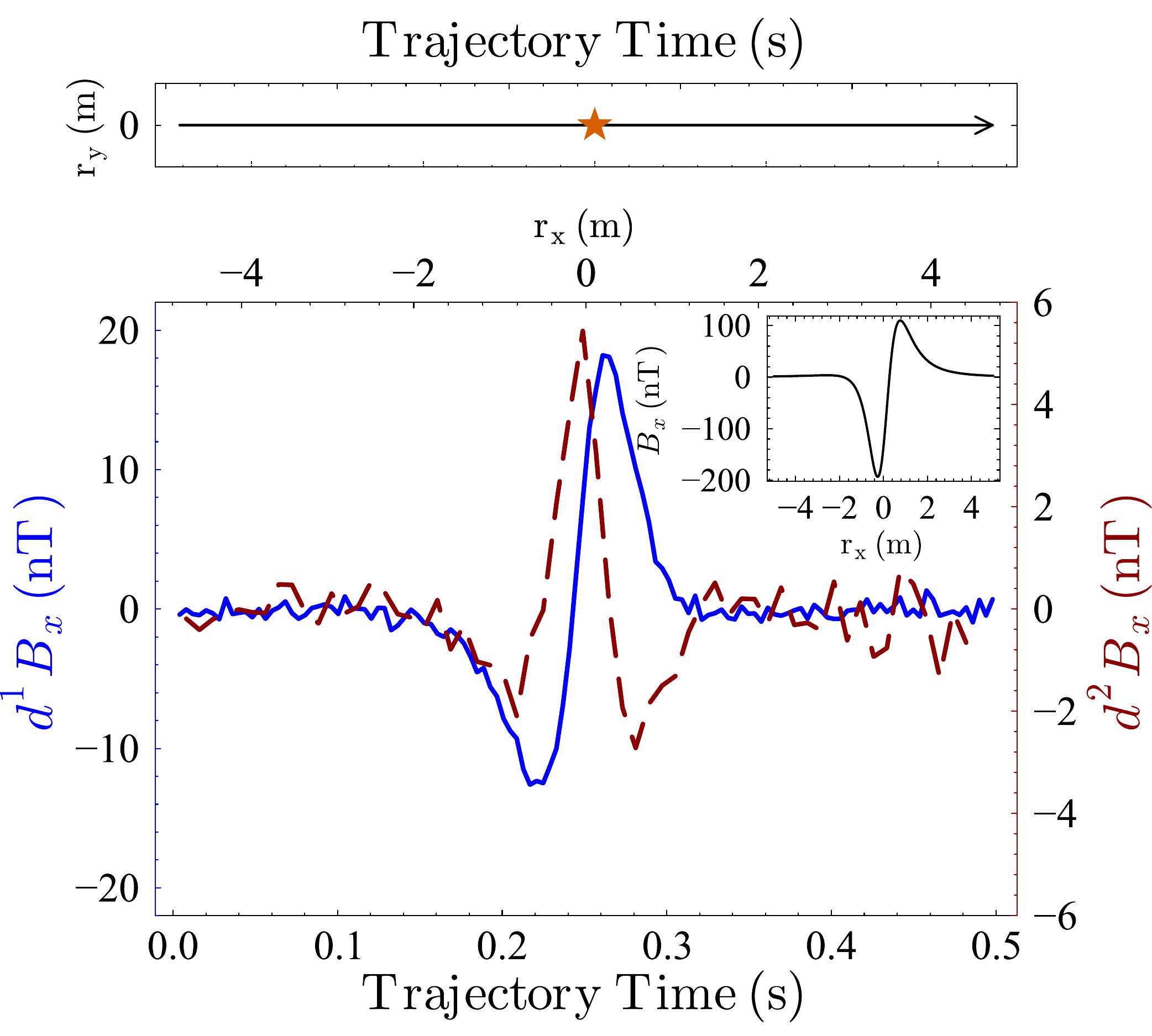}%
    }\hfill
    \subfloat[\label{F:3b}]{%
    \includegraphics[width=0.32\textwidth]{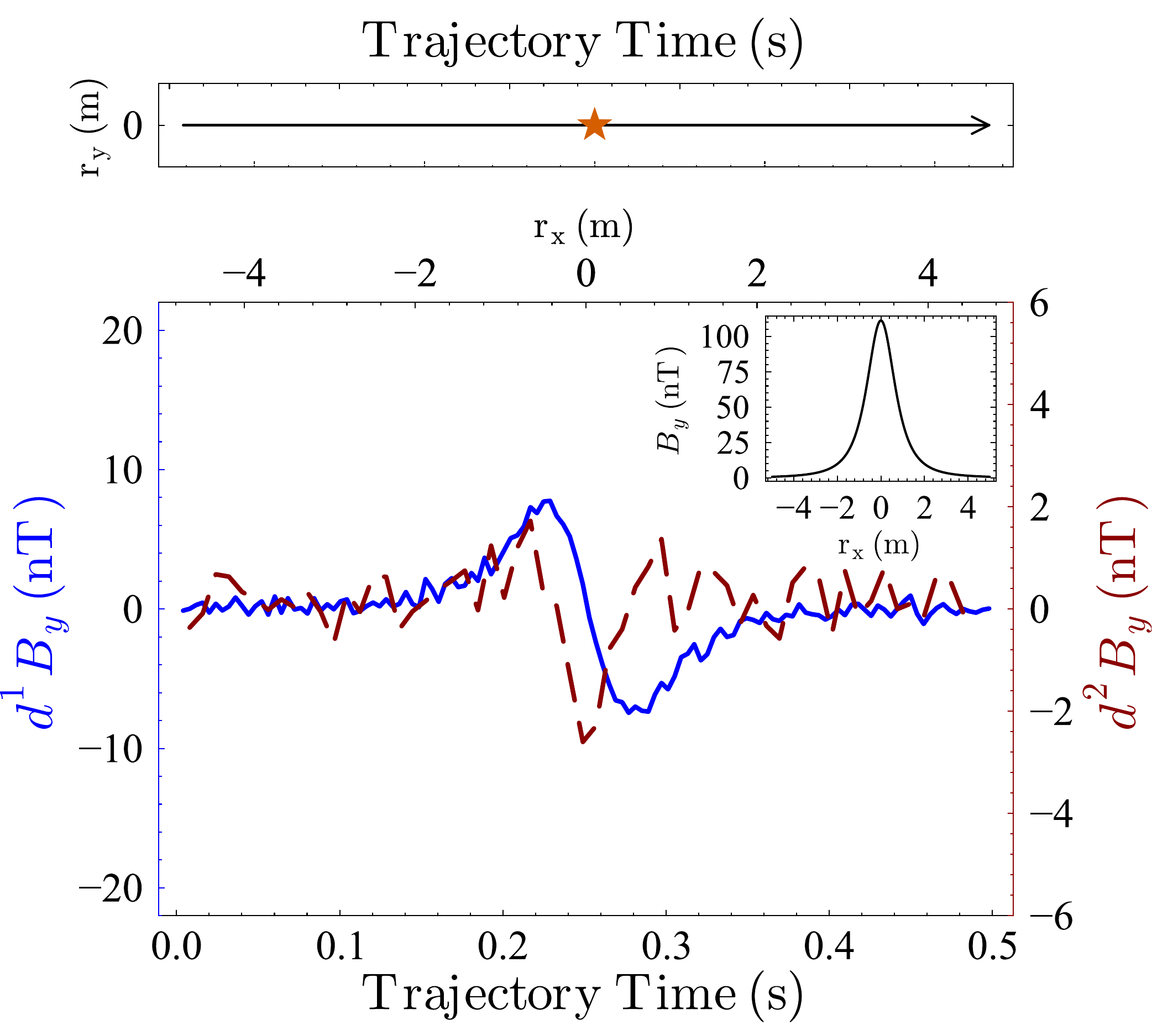}%
    }\hfill
    \subfloat[\label{F:3c}]{%
    \includegraphics[width=0.32\textwidth]{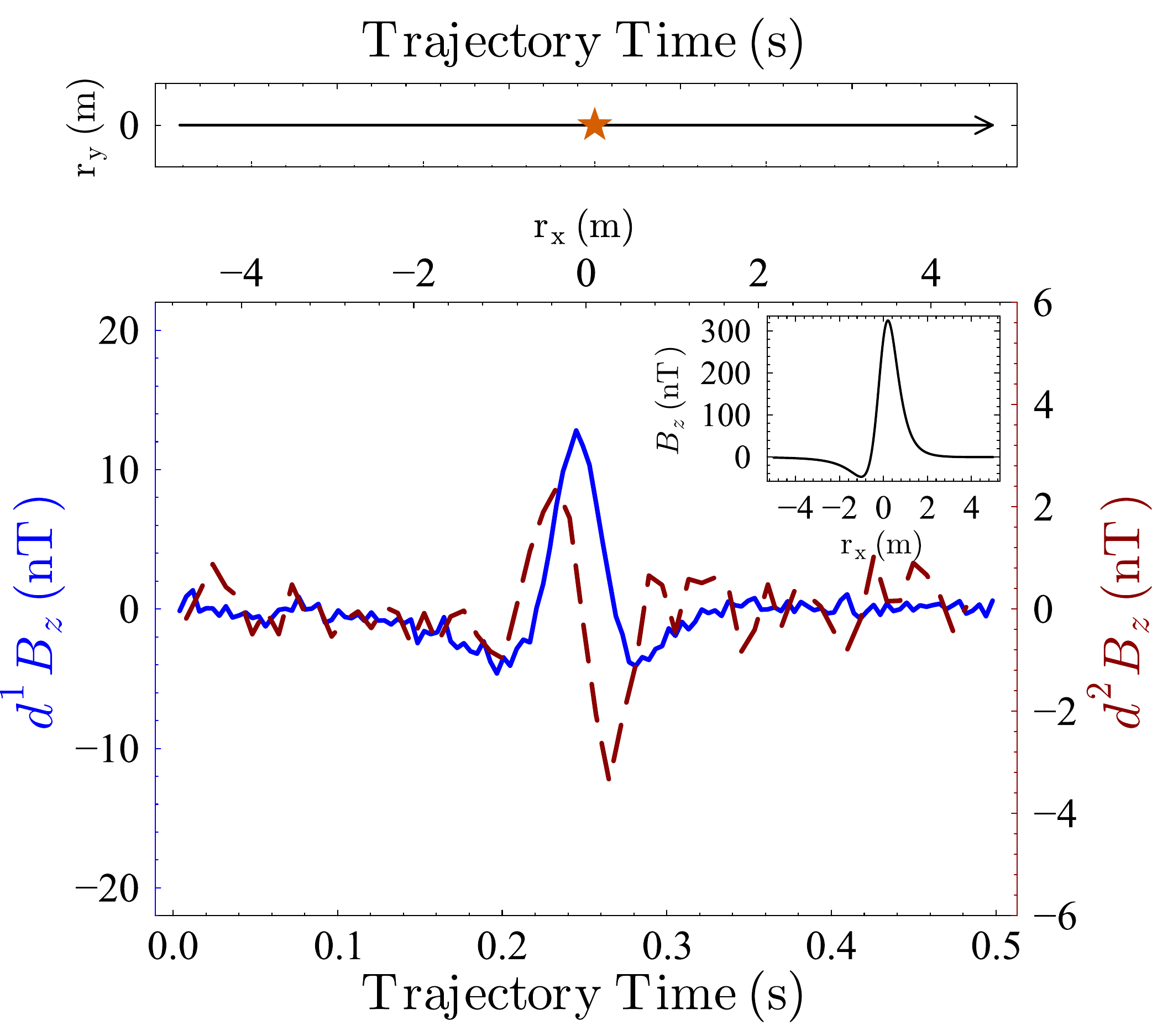}%
    }
    \caption{Numerical simulation of vectorial gradiometry for magnetic anomaly detection. The sensor vibrates at a frequency $\nu/(2\pi) = 1$ MHz and is displaced at a constant speed $v = 20$ m/s and height $h = 1$ m over the source of the magnetic anomaly (marked with a star), which is located at the center of the trajectory (shown in the top panel). The source of the anomaly is a ferromagnetic sphere of radius $R = 16$ cm, which is magnetized by a geomagnetic field of intensity $B = 45.84 \ \mathrm{\mu T}$, colatitude $\vartheta = 105.63^\circ$ and longitude $\varphi = 320.72^\circ$. The intensity of the anomaly field as a function of the sensor position is shown in the inset. Nonetheless, the geomagnetic field is also included in the simulations. Our protocol is applied to measure first ($d^1B$) and second ($d^2B$) order temporal variations (solid and dashed lines, respectively), of all three cartesian coordinates of the field $B_x$, $B_y$ and $B_z$ (panels (a), (b) and (c), respectively), with stage II occuring in the memory state and having a duration $t_{II} = 4$ ms. The geomagnetic field is suppressed in the measurement output. Photon shot noise is accounted for in our simulations, such that the measurement contrast is taken to be $C = 10\%$, the average number of photon collected per NV center is $n_{av} = 0.022$, the number of NVs in the ensemble is $N_{NV} = 10^{11}$ and the number of repetitions is $10^3$. The total duration of the `experiment' is $t = 8.33$ min.}
    \label{F: MagneticAnomaly}
\end{figure*}

We consider an NV-based sensor, where the NV axis (oriented along the $z$ direction in the NV reference frame, $z_{NV}$) forms an angle $\theta_0$ with the $z$ axis in the laboratory reference frame (see Fig. \ref{F: System}). The sensor is mounted on the tip of a cantilever which oscillates with frequency $\nu$ and amplitude $\alpha$, such that the angle between the $z$ and $z_{NV}$ axes becomes time dependent, $\theta_{NV}(t) = \theta_0 + \alpha \sin \nu t$. Hence, the magnetic field detected by the sensor at time $t_i$, $\bm{B}(t_i) = (B_x, B_y, B_z)$, acquires a time dependence in its reference frame, $\bm{B}^{NV}(t) = R_{\hat{k}} (\theta_{NV}(t)) \bm{B}$, where $R_{\hat{k}}(\theta_{NV}(t))$ is a rotation matrix which models the oscillation of the NV sensor and $\hat{k} = (k_x, k_y, k_z)$ is a unit vector perpendicular to the plane of vibration. For the particular case shown in Fig. \ref{F: System}, where $\hat{k} = \hat{x}$, the projection of the time dependent field along the NV axis reads 
\begin{equation}
    B_z^{NV}(t) \approx \alpha \sin\nu t \cos\theta_0 B_y = \mathfrak{B} \sin\nu t,
\end{equation}
where $\mathfrak{B}$ is the effective amplitude of the sensed magnetic field and a small oscillation amplitude is assumed. See Section \ref{SM: DCtoAC} in the Supplementary Information for a more detailed description. 

When combining this oscillatory motion of the sensor with our gradiometry protocol, a dynamical decoupling pulse sequence with an interpulse separation $\tau = 1/(4\nu)$ is applied during stages I and III to capture the imprinted eAC signal. This allows longer durations of the interrogations stages.
\section{Results}\label{S: Results}

In this section, we apply our protocol to two distinct scenarios. First, we discuss the detection of magnetic anomalies and, second, we consider the measurement of magnetic fields originated from neuronal action potentials (APs). For details on the modelling of these systems, see the Supplementary Information Sections \ref{SM: MagAnomalies} and \ref{SM: NeuronAP}. We choose our sensor to be an ensemble of NV centers in diamond comprising $N_{NV} = 10^{11}$ active spins \cite{Wolf_2015,Barry_2016}. Our simulations include photon shot noise modeled as a Gaussian with zero mean and standard deviation $\sigma_R = 1/(C\sqrt{n_{\mathrm{av}}N_{NV}})$ \cite{Barry_2020}, such that a fluorescence contrast $C = 10\%$ and an average number of photons received per spin $n_\mathrm{{av}} = 0.022$ are assumed, and $10^3$ `experimental' repetitions are realized.

\subsection{Magnetic Anomaly Detetction}

When placed in a magnetic field, a ferromagnetic object magnetizes, giving rise to a secondary field which oposes the first. In geological applications, the magnetizing field is the geomagnetic field ($B_{gm} \sim \mathrm{\mu T}$), while the secondary is called magnetic anomaly ($B_a \sim \mathrm{nT}$), which is typically originated from underground ferromagnetic sources. The detection and characterization of magnetic anomalies is relevant to distinct applications in geophysical research \cite{Dang_2010,Gang_2016,Newman_2024,Levi_2025}.

Often, magnetic anomalies show spatial variations on a shorter scale than the geomagnetic field, such that the latter can be considered static across the anomaly. By displacing the sensor at a constant speed $v$, the magnetic anomaly acquires a time dependence, in the frame of the sensor, which fits within the methodology of our protocol. We consider our source of magnetic anomaly to be a ferromagnetic sphere of radius $R = 16$ cm and a geomagnetic field of intensity $B = 45.85 \mathrm{\mu T}$, colatitude $\vartheta = 105.63^\circ$ and longitude $\varphi = 320.72^\circ$ (see Section \ref{SM: MagAnomalies} in the Supplementary Information for details on the modelling of the magnetic anomaly). 

Figure \ref{F: MagneticAnomaly} shows the numerical results obtained for an NV ensemble which vibrates at a frequency $\nu/(2\pi) = 1$ MHz, and is displaced at a constant speed $v = 20\ \mathrm{m/s}$ and height $h = 1$ m following a straight trajectory over the spheric source. The inset in Fig. \ref{F: MagneticAnomaly} shows the intensity of the magnetic anomaly as a function of the distance to the source, which is located at the center of the trajectory (\textit{i.e.}, ar $r_x = 0$). The evolution during stage II occurs in the state of the nitrogen nucleus, such that the duration of this stage is $t_{II} = 4$ ms. A CPMG dynamical decoupling sequence with an interpulse separation $\tau = 1/(4\nu)$ is applied to the sensor during stages I and III to capture a phase due to the effective AC signal, which reads
\begin{equation}
    \phi = \frac{2 N_p \alpha |\gamma_e| B_i}{\nu},
\end{equation}
where $\gamma_e = -28.024 \mathrm{GHz/T}$ is the electron gyromagnetic ratio, $N_p = 16$ is the number of pulses in the CPMG sequence and $B_i$ is some component of the total field in the laboratory reference frame. If this is taken such that $\theta_0 = 0$ (\textit{i.e.}, the NV is aligned with the $z$-axis in the laboratory frame) and $\hat{k} = \hat{x}$ ($\hat{k} = \hat{y}$), then $B_i = B_y$ ($B_i = -B_x$). If, however, $\theta_0 = \pi/2$ and $\hat{k}$ lies in the $xy$ plane, then $B_i = -B_z$. Therefore, by considering these three different geometric configurations of the system, vector gradiometry is enabled (see Section \ref{SM: DCtoAC} in the Supplementary Information for more details). 

We apply our gradiometry protocol to measure first and second order variations of the field , $d^1B$ and $d^2B$ (in blue and red, respectively), and show that in both cases the static geomagnetic field is suppressed. Note that the $n$-th order variation of the field is given by $d^{(n)}B = \frac{\nu}{2N_p\alpha|\gamma_e|} \langle \sigma_z \rangle^{(n)}$. We show results for the three cartesian components of the field, showcasing the applicability of our protocol for vector gradiometry. 

\subsection{Magnetophysiology from neuron action potentials}

For biosensing applications, magnetophysiology has gained huge interest in recent years \cite{Webb_2021,Zhang_2023,Klein_2023,Wu_2025,Omar_2026}. While electrophysiology is easier to perform due to the stronger electric signals, magnetic biosignals are less affected by biological tissue and also carry directional information about the field \cite{Klein_2023}. Highly sensitive quantum sensors open the door to magnetic biosensing, with many applications in life sciences, such as magnetocardiography \cite{Arai_2022,Yu_2024,Omar_2026} and magnetoencephalography \cite{Barry_2016,Masuyama_2021,Alem_2023}. In particular, monitoring neuronal activity from single cells is relevant for early detection of neurodegenerative diseases. However, magnetic biosensing requires shielding from the environment, which is expensive and often weak, resulting in an imperfect suppression of the external noise. 

Here, we study the magnetic field associated with an electrical impulse in a neuron axon. Modelling the axon as a cylinder, the Biot-Savart law gives the corresponding magnetic field, $B(t) = s\frac{d\Phi}{dt}$, where $\Phi$ is the intracellular voltage due to the movement of ions in the neuron, which gives rise to the electrical impulse, and $s$ is a proportionality constant that depends on geometrical and physiological quantities (see Section \ref{SM: NeuronAP} in the Supplementary Information for more details about the model). Following Ref. \cite{Barry_2016}, where the authors study magnetic fields from neuronal APs in a giant axon from a marine fanworm, we take $s = 7.6 \mathrm{pT / (V s^{-1})}$. 

\begin{figure}[t]
    \centering
    \includegraphics[width=0.45\textwidth]{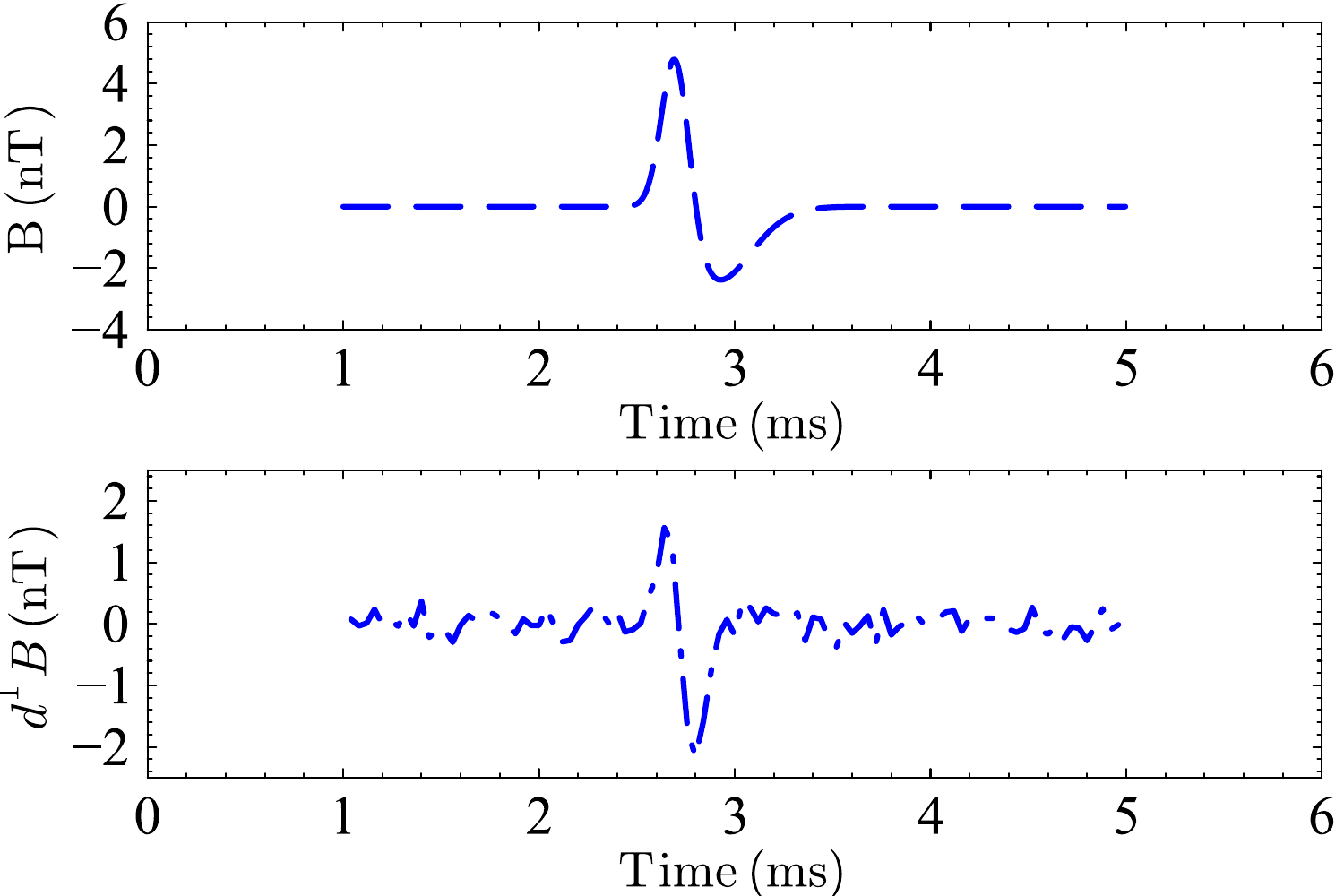}
    \caption{(Upper panel) Expected magnetic field from a neuron AP in a giant axon from a marine worm. (Lower panel) Differential signal measured from our gradiometry protocol. Stage II occurs in the sensor state and has a duration of $32 \ \mathrm{\mu s}$. Photon shot noise is accounted for in our simulation, such that the measurement contrast is $C = 10\%$, the average number of photons received per NV center is $n_{av} = 0.022$, the number of NVs in the ensemble is $N_{NV} = 10^{11}$ and the number of repetitions is $10^3$. Note that the $x$-axis corresponds to a single run, such that the total duration of the `experiment' is $t = 4$s.}
    \label{F: NeuronAP}
\end{figure}

The sensor setup is considered the same as in the previous example, although in this case the sensor is not moving, and the evolution in stage II occurs in the sensor state and has a duration $t_{II} = 32 \ \mathrm{\mu s}$. The magnetic field originated from the neuron AP is modeled to replicate the experimental measurement of Ref. \cite{Barry_2016} and is depicted in Fig. \ref{F: NeuronAP} (upper panel). The lower panel in the figure shows the differential signal obtained from our gradiometry protocol. Our results show that a differential measurement of magnetic fields originated from neuron action potentials could be performed, thus not requiring magnetic shielding from the environment. Remarkably, because our protocol can be applied for vectorial magnetometry, we could recover the directional information of the magnetic field, which is one of the main advantages of magnetophysiology over electrophysiology \cite{Klein_2023}.

\section{Discussion}\label{S: Discussion}

It is important to note that, if the differential signal is taken between two points with a small temporal or spatial separation between them, the difference between the field values may be close to zero, such that the signal amplitude is heavily reduced, or even suppressed. In order to enlarge this difference, it is possible to extend the duration of the displacement stage, which is limited by the dephasing time of the sensor, $T_2^*$. However, because the phase accumulated during this stage has to be small, dynamical decoupling pulse sequences may be applied such that the stage is limited by the sensor coherence time $T_2$. Nonetheless, care must be taken in choosing the number of pulses, as an odd multiple causes the phases in Eq.(\ref{Eq. Gradiometry}) to add rather than subtract, thereby compounding the effect of any background field.

Additionally, the design of the sequence allows for the inclusion of driving fields which suppress unfavorable dipolar interactions between spins \cite{Biteri_2025,Olivan_2025}, further increasing the coherence time of the sensor without damaging the functionality of the protocol. Moreover, as demonstrated in the previous sections, the cohererence time can be even further extended to the nuclear $T_2$.

Lastly, for the particular case where the target field shows a spatial dependence, physical motion of the sensor enables vectorial gradiometry of the field along the direction of movement. Thus, by changing the direction of motion of the sensor, tensor gradiometry can be performed, which results in a more robust characterization of the field, as it is less orientation dependent.

\textit{Data availability}--- The data supporting this study are not publicly available. The data are available from the authors upon reasonable request.

\begin{acknowledgements}
Authors acknowledge support by the European Union's Horizon Europe research and innovation programme under Grant Agreement No. 101135742 (QUENCH). J. C. acknowledges the Agencia Estatal de Investigaci\'{o}n via the Modelizado, Optimizaci\'{o}n, y Esquemas de Magnetometria en Centros de Color project PID2024-161371NB-C22, and the Basque Government under Grant No. IT1470-22.  
\end{acknowledgements}

\bibliography{lib}

@article{Dang_2010,
    author = {Dang, H. B. and Maloof, A. C. and Romalis, M. V.},
    title = {Ultrahigh sensitivity magnetic field and magnetization measurements with an atomic magnetometer},
    journal = {Applied Physics Letters},
    volume = {97},
    number = {15},
    pages = {151110},
    year = {2010},
    month = {10},
    issn = {0003-6951},
    doi = {10.1063/1.3491215},
    url = {https://doi.org/10.1063/1.3491215},
}

@article{Gang_2016,
title = {Detection of ferromagnetic target based on mobile magnetic gradient tensor system},
journal = {Journal of Magnetism and Magnetic Materials},
volume = {402},
pages = {1-7},
year = {2016},
issn = {0304-8853},
doi = {https://doi.org/10.1016/j.jmmm.2015.11.034},
url = {https://www.sciencedirect.com/science/article/pii/S0304885315308064},
author = {Y.I.N. Gang and Zhang Yingtang and Li Zhining and Fan Hongbo and Ren Guoquan},
}

@article{Newman_2024,
  title = {Tensor gradiometry with a diamond magnetometer},
  author = {Newman, A.J. and Graham, S.M. and Edmonds, A.M. and Twitchen, D.J. and Markham, M.L. and Morley, G.W.},
  journal = {Phys. Rev. Appl.},
  volume = {21},
  issue = {1},
  pages = {014003},
  numpages = {15},
  year = {2024},
  month = {Jan},
  publisher = {American Physical Society},
  doi = {10.1103/PhysRevApplied.21.014003},
  url = {https://link.aps.org/doi/10.1103/PhysRevApplied.21.014003}
}

@article{Levi_2025,
author = {Kfir Levi and Avital Giat and Liran Golan and Eliran Talker and Liron Stern},
journal = {Optica Quantum},
number = {1},
pages = {84--92},
publisher = {Optica Publishing Group},
title = {Remote chip-scale quantum sensing of magnetic fields},
volume = {3},
month = {Feb},
year = {2025},
url = {https://opg.optica.org/opticaq/abstract.cfm?URI=opticaq-3-1-84},
doi = {10.1364/OPTICAQ.528399},
}

@article{Huxter_2022,
author = {Huxter, W. S. and Palm, M. L. and Davis, M. L. and Welter, P. and Lambert, C. -H. and Trassin, M. and Degen, C. L.},
title = {Scanning gradiometry with a single spin quantum magnetometer},
journal = {Nature Communications},
volume = {13},
number = {1},
pages = {3761},
year = {2022},
doi = {10.1038/s41467-022-31454-6},
URL = {https://doi.org/10.1038/s41467-022-31454-6}
}

@article{Huxter_2023,
author = {Huxter, William S. and Sarott, Martin F. and Trassin, Morgan and Degen, Christian L.},
title = {Imaging ferroelectric domains with a single-spin scanning quantum sensor},
journal = {Nature Physics},
volume = {19},
number = {5},
pages = {644--648},
year = {2023},
doi = {10.1038/s41567-022-01921-4},
URL = {https://doi.org/10.1038/s41567-022-01921-4}
}

@article{Melendez_2025,
author = {Alex L. Melendez  and Shekhar Das  and Francisco Ayala Rodriguez  and I-Hsuan Kao  and Wenhao Liu  and Archibald J. Williams  and Bing Lv  and Joshua Goldberger  and Shubhayu Chatterjee  and Simranjeet Singh  and P. Chris Hammel },
title = {Quantum sensing of broadband spin dynamics and magnon transport in antiferromagnets},
journal = {Science Advances},
volume = {11},
number = {26},
pages = {eadu9381},
year = {2025},
doi = {10.1126/sciadv.adu9381},
}

@article{Barry_2016,
author = {John F. Barry  and Matthew J. Turner  and Jennifer M. Schloss  and David R. Glenn  and Yuyu Song  and Mikhail D. Lukin  and Hongkun Park  and Ronald L. Walsworth },
title = {Optical magnetic detection of single-neuron action potentials using quantum defects in diamond},
journal = {Proceedings of the National Academy of Sciences},
volume = {113},
number = {49},
pages = {14133-14138},
year = {2016},
doi = {10.1073/pnas.1601513113},
}

@article{Webb_2021,
  title = {Detection of biological signals from a live mammalian muscle using an early stage diamond quantum sensor},
  author = {Webb, James Luke and Troise, Luca and Hansen, Nikolaj Winther and Olsson, Christoffer and Wojciechowski, Adam M. and Achard, Jocelyn and Brinza, Ovidiu and Staacke, Robert and Kieschnick, Michael and Meijer, Jan and Thielscher, Axel and Perrier, Jean-François and Berg-Sørensen, Kirstine and Huck, Alexander and Andersen, Ulrik Lund},
  journal = {Scientific Reports},
  volume = {11},
  issue = {1},
  pages = {2412},
  year = {2021},
  doi = {10.1038/s41598-021-81828-x},
  url = {https://doi.org/10.1038/s41598-021-81828-x}
}

@article{Arai_2022,
author = {Arai, Keigo and Kuwahata, Akihiro and Nishitani, Daisuke and Fujisaki, Ikuya and Matsuki, Ryoma and Nishio, Yuki and Xin, Zonghao and Cao, Xinyu and Hatano, Yuji and Onoda, Shinobu and Shinei, Chikara and Miyakawa, Masashi and Taniguchi, Takashi and Yamazaki, Masatoshi and Teraji, Tokuyuki and Ohshima, Takeshi and Hatano, Mutsuko and Sekino, Masaki and Iwasaki, Takayuki},
title = {Millimetre-scale magnetocardiography of living rats with thoracotomy},
journal = {Communications Physics},
volume = {5},
number = {1},
pages = {200},
year = {2022},
doi = {10.1038/s42005-022-00978-0},
URL = {https://doi.org/10.1038/s42005-022-00978-0}
}

@article{Yu_2024,
  title = {Noninvasive magnetocardiography of a living rat based on a diamond quantum sensor},
  author = {Yu, Ziyun and Xie, Yijin and Jin, Guodong and Zhu, Yunbin and Zhang, Qi and Shi, Fazhan and Wan, Fang-yan and Luo, Hongmei and Tang, Ai-hui and Rong, Xing},
  journal = {Phys. Rev. Appl.},
  volume = {21},
  issue = {6},
  pages = {064028},
  numpages = {12},
  year = {2024},
  month = {Jun},
  publisher = {American Physical Society},
  doi = {10.1103/PhysRevApplied.21.064028},
  url = {https://link.aps.org/doi/10.1103/PhysRevApplied.21.064028}
}

@article{Chwala_2019,
doi = {10.1088/1361-6668/aaf245},
url = {https://doi.org/10.1088/1361-6668/aaf245},
year = {2019},
month = {jan},
publisher = {IOP Publishing},
volume = {32},
number = {2},
pages = {024003},
author = {Chwala, A and Schmelz, M and Zakosarenko, V and Schiffler, M and Schneider, M and Thürk, M and Bräuer, S and Bauer, F and Schulz, M and Krüger, A and Stolz, R},
title = {Underwater operation of a full tensor SQUID gradiometer system},
journal = {Superconductor Science and Technology},
}

@article{Gkika_2024,
doi = {10.1007/s10909-024-03152-8},
url = {https://doi.org/10.1007/s10909-024-03152-8},
year = {2024},
publisher = {IOP Publishing},
volume = {216},
number = {1},
pages = {386--396},
author = {Gkika, Violeta and Kim, Younggeun and Matlashov, Andrei and Shin, Yun Chang and Semertzidis, Yannis and Cantor, Robin and Lohmeyer, Chloe and Aggarwal, Nancy and Geraci, Andrew},
title = {Optimization of High-Sensitivity SQUID Gradiometer for ARIADNE at CAPP},
journal = {Journal of Low Temperature Physics},
}

@article{Sheng_2017,
    author = {Sheng, D. and Perry, A. R. and Krzyzewski, S. P. and Geller, S. and Kitching, J. and Knappe, S.},
    title = {A microfabricated optically-pumped magnetic gradiometer},
    journal = {Applied Physics Letters},
    volume = {110},
    number = {3},
    pages = {031106},
    year = {2017},
    month = {01},
    issn = {0003-6951},
    doi = {10.1063/1.4974349},
    url = {https://doi.org/10.1063/1.4974349},
}

@article{Cook_2024,
doi = {10.1088/2058-9565/ad3d81},
url = {https://doi.org/10.1088/2058-9565/ad3d81},
year = {2024},
month = {apr},
publisher = {IOP Publishing},
volume = {9},
number = {3},
pages = {035016},
author = {Cook, Harry and Bezsudnova, Yulia and Koponen, Lari M and Jensen, Ole and Barontini, Giovanni and Kowalczyk, Anna U},
title = {An optically pumped magnetic gradiometer for the detection of human biomagnetism},
journal = {Quantum Science and Technology},
}

@article{Blakley_2018,
    author = {Blakley, S. M. and Fedotov, I. V. and Becker, J. and Zheltikov, A. M.},
    title = {Quantum stereomagnetometry with a dual-core photonic-crystal fiber},
    journal = {Applied Physics Letters},
    volume = {113},
    number = {1},
    pages = {011112},
    year = {2018},
    month = {07},
    issn = {0003-6951},
    doi = {10.1063/1.5024583},
    url = {https://doi.org/10.1063/1.5024583},
}

@article{Masuyama_2021,
AUTHOR = {Masuyama, Yuta and Suzuki, Katsumi and Hekizono, Akira and Iwanami, Mitsuyasu and Hatano, Mutsuko and Iwasaki, Takayuki and Ohshima, Takeshi},
TITLE = {Gradiometer Using Separated Diamond Quantum Magnetometers},
JOURNAL = {Sensors},
VOLUME = {21},
YEAR = {2021},
NUMBER = {3},
ARTICLE-NUMBER = {977},
URL = {https://www.mdpi.com/1424-8220/21/3/977},
PubMedID = {33540515},
ISSN = {1424-8220},
}

@article{Wood_2021_2,
  title = {${T}_{2}$-limited sensing of static magnetic fields via fast rotation of quantum spins},
  author = {Wood, A. A. and Aeppli, A. G. and Lilette, E. and Fein, Y. Y. and Stacey, A. and Hollenberg, L. C. L. and Scholten, R. E. and Martin, A. M.},
  journal = {Phys. Rev. B},
  volume = {98},
  issue = {17},
  pages = {174114},
  numpages = {8},
  year = {2018},
  month = {Nov},
  publisher = {American Physical Society},
  doi = {10.1103/PhysRevB.98.174114},
  url = {https://link.aps.org/doi/10.1103/PhysRevB.98.174114}
}

@article{Wood_2022,
  title = {dc Quantum Magnetometry below the Ramsey Limit},
  author = {Wood, Alexander A. and Stacey, Alastair and Martin, Andy M.},
  journal = {Phys. Rev. Appl.},
  volume = {18},
  issue = {5},
  pages = {054019},
  numpages = {13},
  year = {2022},
  month = {Nov},
  publisher = {American Physical Society},
  doi = {10.1103/PhysRevApplied.18.054019},
  url = {https://link.aps.org/doi/10.1103/PhysRevApplied.18.054019}
}

@article{Zhang_2020,
  title = {Portable intrinsic gradiometer for ultra-sensitive detection of magnetic gradient in unshielded environment},
  author = {Zhang, Rui and Mhaskar, Rahul and Smith, Ken and Prouty, Mark},
  journal = {Applied Physics Letters},
  volume = {116},
  issue = {14},
  pages = {143501},
  year = {2020},
  doi = {10.1063/5.0004746},
  url = {https://doi.org/10.1063/5.0004746}
}

@article{Zhang_2016,
  author={Zhang, Rui and Smith, Kenneth and Mhaskar, Rahul},
  journal={Proc. IEEE SENSORS},
  title={Highly sensitive miniature scalar optical gradiometer}, 
  year={2016},
  pages={1--3},
  doi={10.1109/ICSENS.2016.7808768},
  url={https://doi.org/10.1109/ICSENS.2016.7808768}}

@article{Zilinski_2026,
  AUTHOR = {Zilinski, Nicholaus and Parameswaran, Ash M. and Gray, Bonnie L. and Cheung, Teresa},
  TITLE = {A Single-Cell Optically Pumped Intrinsic Gradiometer},
  JOURNAL = {Sensors},
  VOLUME = {26},
  YEAR = {2026},
  NUMBER = {5},
  ARTICLE-NUMBER = {1678},
  URL = {https://www.mdpi.com/1424-8220/26/5/1678},
  }

@article{Wood_2018,
author = {Alexander A. Wood  and Emmanuel Lilette  and Yaakov Y. Fein  and Nikolas Tomek  and Liam P. McGuinness  and Lloyd C. L. Hollenberg  and Robert E. Scholten  and Andy M. Martin },
title = {Quantum measurement of a rapidly rotating spin qubit in diamond},
journal = {Science Advances},
volume = {4},
number = {5},
year = {2018},
URL = {https://www.science.org/doi/abs/10.1126/sciadv.aar7691},
}

@article{Wood_2021_1,
  title = {Quantum control of nuclear-spin qubits in a rapidly rotating diamond},
  author = {Wood, A. A. and Goldblatt, R. M. and Scholten, R. E. and Martin, A. M.},
  journal = {Phys. Rev. Res.},
  volume = {3},
  issue = {4},
  pages = {043174},
  numpages = {12},
  year = {2021},
  month = {Dec},
  publisher = {American Physical Society},
  doi = {10.1103/PhysRevResearch.3.043174},
  url = {https://link.aps.org/doi/10.1103/PhysRevResearch.3.043174}
}

@article{Liu_2019,
  title = {Nanoscale Vector dc Magnetometry via Ancilla-Assisted Frequency Up-Conversion},
  author = {Liu, Yi-Xiang and Ajoy, Ashok and Cappellaro, Paola},
  journal = {Phys. Rev. Lett.},
  volume = {122},
  issue = {10},
  pages = {100501},
  numpages = {6},
  year = {2019},
  month = {Mar},
  publisher = {American Physical Society},
  doi = {10.1103/PhysRevLett.122.100501},
  url = {https://link.aps.org/doi/10.1103/PhysRevLett.122.100501}
}

@misc{Richter_2026,
      title={Robust Quantum Sensing via Prethermal Spin Orbits}, 
      author={Enrico Daniel Richter and Ryan J. Smith and Brayden Glockzin and Emanuel Druga and Thomas Schenkel and Ashok Ajoy},
      year={2026},
      eprint={2603.21057},
      archivePrefix={arXiv},
      primaryClass={quant-ph},
}

@article{Voce_2025,
  title = {Efficient radio-frequency sensing with fluorescence encoding},
  author = {Voce, N. and Stevenson, P.},
  journal = {Phys. Rev. Appl.},
  volume = {26},
  issue = {1},
  pages = {014068},
  numpages = {11},
  year = {2026},
  month = {Jul},
  publisher = {American Physical Society},
  doi = {10.1103/pk7l-f998},
  url = {https://link.aps.org/doi/10.1103/pk7l-f998}
}

@article{Taylor_2008,
  title = {High-sensitivity diamond magnetometer with nanoscale resolution},
  author = {Taylor, J. M. and Cappellaro, P. and Childress, L. and Jiang, L. and Budker, D. and Hemmer, P. R. and Yacoby, A. and Walsworth, R. and Lukin, M. D.},
  journal = {Nature Physics},
  volume = {4},
  issue = {10},
  pages = {810--816},
  year = {2008},
  doi = {10.1038/nphys1075},
  url = {https://doi.org/10.1038/nphys1075}
}

@article{Dreau_2011,
  title = {Avoiding power broadening in optically detected magnetic resonance of single NV defects for enhanced dc magnetic field sensitivity},
  author = {Dr\'eau, A. and Lesik, M. and Rondin, L. and Spinicelli, P. and Arcizet, O. and Roch, J.-F. and Jacques, V.},
  journal = {Phys. Rev. B},
  volume = {84},
  issue = {19},
  pages = {195204},
  numpages = {8},
  year = {2011},
  month = {Nov},
  publisher = {American Physical Society},
  doi = {10.1103/PhysRevB.84.195204},
  url = {https://link.aps.org/doi/10.1103/PhysRevB.84.195204}
}

@article{Wu_2025,
doi = {10.1088/1361-6528/adb635},
url = {https://doi.org/10.1088/1361-6528/adb635},
year = {2025},
month = {feb},
publisher = {IOP Publishing},
volume = {36},
number = {15},
pages = {152501},
author = {Wu, Kai and He, Rui},
title = {Perspective: magnetic quantum sensors for biomedical applications},
journal = {Nanotechnology},
}

@article{Klein_2023,
author = {Klein, Frederike J. and Jendritza, Patrick and Chopin, Chlo{\'e} and Parto-Dezfouli, Mohsen and Solignac, Aur{\'e}lie and Fermon, Claude and Pannetier-Lecoeur, Myriam and Fries, Pascal},
title = {In vivo magnetic recording of single-neuron action potentials},
journal = {Journal of Neurophysiology},
volume = {134},
number = {4},
pages = {1306-1319},
year = {2025},
doi = {10.1152/jn.00491.2024},
URL = {https://doi.org/10.1152/jn.00491.2024}
}

@ARTICLE{Zhang_2023,
    
AUTHOR={Zhang, Chen  and Zhang, Jixing  and Widmann, Matthias  and Benke, Magnus  and Kübler, Michael  and Dasari, Durga  and Klotz, Thomas  and Gizzi, Leonardo  and Röhrle, Oliver  and Brenner, Philipp  and Wrachtrup, Jörg },
TITLE={Optimizing NV magnetometry for Magnetoneurography and Magnetomyography applications},
JOURNAL={Frontiers in Neuroscience},
VOLUME={Volume 16},
YEAR={2023},
URL={https://www.frontiersin.org/journals/neuroscience/articles/10.3389/fnins.2022.1034391},
}

@ARTICLE{Alem_2023,
    
AUTHOR={Alem, Orang  and Hughes, K. Jeramy  and Buard, Isabelle  and Cheung, Teresa P.  and Maydew, Tyler  and Griesshammer, Andreas  and Holloway, Kendall  and Park, Aaron  and Lechuga, Vanessa  and Coolidge, Collin  and Gerginov, Marja  and Quigg, Erik  and Seames, Alexander  and Kronberg, Eugene  and Teale, Peter  and Knappe, Svenja },
TITLE={An integrated full-head OPM-MEG system based on 128 zero-field sensors},
JOURNAL={Frontiers in Neuroscience},
VOLUME={Volume 17},
YEAR={2023},
URL={https://www.frontiersin.org/journals/neuroscience/articles/10.3389/fnins.2023.1190310},
}

@misc{Olivan_2025,
      title={Suppressing Fast Dipolar Noise in Solid-State Spin Qubits}, 
      author={García Oliván, Jaime and Biteri-Uribarren, Ainitze and Whaites, Oliver T. and Casanova, Jorge},
      year={2025},
      eprint={2512.06948},
      archivePrefix={arXiv},
      primaryClass={quant-ph},
}

@misc{Xie_2022,
      title={${T}_{2}$-limited dc Quantum Magnetometry via Flux Modulation}, 
      author={Yijin Xie and Caijin Xie and Yunbin Zhu and Ke Jing and Yu Tong and Xi Qin and Haosen Guan and Chang-Kui Duan and Ya Wang and Xing Rong and Jiangfeng Du},
      year={2022},
      eprint={2204.07343},
      archivePrefix={arXiv},
      primaryClass={quant-ph},
}

@article{Alsina_2025,
  title = {Enhanced microscale NMR spectroscopy of low-gyromagnetic ratio nuclei via hydrogen transfer},
  author = {Alsina-Bol\'{\i}var, P. and Casanova, J.},
  journal = {Phys. Rev. Res.},
  volume = {7},
  issue = {2},
  pages = {023258},
  numpages = {11},
  year = {2025},
  month = {Jun},
  publisher = {American Physical Society},
  doi = {10.1103/PhysRevResearch.7.023258},
  url = {https://link.aps.org/doi/10.1103/PhysRevResearch.7.023258}
}

@misc{Grafenstein_2025,
      title={Coherent signal detection in the statistical polarization regime enables high-resolution nanoscale NMR spectroscopy}, 
      author={Nick R. von Grafenstein and Karl D. Briegel and Jorge Casanova and Dominik B. Bucher},
      year={2025},
      eprint={2501.02093},
      archivePrefix={arXiv},
      primaryClass={physics.chem-ph},
}

@article{Katsumi_2025,
      title={High-sensitivity nanoscale quantum sensors based on a diamond micro-resonator}, 
      author={Katsumi, Ryota and Takada, Kosuke and Kawai, Kenta and Sato, Daichi and Yatsui, Takashi},
      journal = {Communications Materials},
      volume = {6},
      issue = {1},
      numpages = {49},
      year={2025},
      url = {https://doi.org/10.1038/s43246-025-00770-x}
}

@article{Whaites_2026,
  title = {Enhanced sensitivity in microscale high-field NMR via nuclear-spin locking with N-V centers},
  author = {Whaites, Oliver T. and Garc\'{\i}a Oliv\'an, Jaime and Casanova, Jorge},
  journal = {Phys. Rev. Appl.},
  volume = {25},
  issue = {1},
  pages = {014062},
  numpages = {17},
  year = {2026},
  month = {Jan},
  publisher = {American Physical Society},
  doi = {10.1103/m4vq-rgrn},
  url = {https://link.aps.org/doi/10.1103/m4vq-rgrn}
}

@article{Schloss_2018,
  title = {Simultaneous Broadband Vector Magnetometry Using Solid-State Spins},
  author = {Schloss, Jennifer M. and Barry, John F. and Turner, Matthew J. and Walsworth, Ronald L.},
  journal = {Phys. Rev. Appl.},
  volume = {10},
  issue = {3},
  pages = {034044},
  numpages = {16},
  year = {2018},
  month = {Sep},
  publisher = {American Physical Society},
  doi = {10.1103/PhysRevApplied.10.034044},
  url = {https://link.aps.org/doi/10.1103/PhysRevApplied.10.034044}
}

@article{Wolf_2015,
  title = {Subpicotesla Diamond Magnetometry},
  author = {Wolf, Thomas and Neumann, Philipp and Nakamura, Kazuo and Sumiya, Hitoshi and Ohshima, Takeshi and Isoya, Junichi and Wrachtrup, J\"org},
  journal = {Phys. Rev. X},
  volume = {5},
  issue = {4},
  pages = {041001},
  numpages = {10},
  year = {2015},
  month = {Oct},
  publisher = {American Physical Society},
  doi = {10.1103/PhysRevX.5.041001},
  url = {https://link.aps.org/doi/10.1103/PhysRevX.5.041001}
}

@article{Barry_2020,
  title = {Sensitivity optimization for NV-diamond magnetometry},
  author = {Barry, John F. and Schloss, Jennifer M. and Bauch, Erik and Turner, Matthew J. and Hart, Connor A. and Pham, Linh M. and Walsworth, Ronald L.},
  journal = {Rev. Mod. Phys.},
  volume = {92},
  issue = {1},
  pages = {015004},
  numpages = {68},
  year = {2020},
  month = {Mar},
  publisher = {American Physical Society},
  doi = {10.1103/RevModPhys.92.015004},
  url = {https://link.aps.org/doi/10.1103/RevModPhys.92.015004}
}

@misc{Omar_2026,
      title={Human Cardiac Measurements with Diamond Magnetometers}, 
      author={Muhib Omar and Magnus Benke and Shaowen Zhang and Jixing Zhang and Michael Kuebler and Pouya Sharbati and Ara Rahimpour and Arno Gueck and Maryna Kapitonova and Devyani Kadam and Carlos Rene Izquierdo Geiser and Jens Haller and Arno Trautmann and Katharina Jag-Lauber and Robert Roelver and Thanh-Duc Nguyen and Leonardo Gizzi and Michelle Schweizer and Mena Abdelsayed and Ingo Wickenbrock and Andrew M. Edmonds and Matthew Markham and Peter A. Koss and Oliver Schnell and Ulrich G. Hofmann and Tonio Ball and Juergen Beck and Dmitry Budker and Joerg Wrachtrup and Arne Wickenbrock},
      year={2026},
      eprint={2601.18843},
      archivePrefix={arXiv},
      primaryClass={physics.med-ph},
}

@article{Biteri_2025,
  title = {Microscale sensing with strongly interacting NV ensembles at high fields},
  author = {Biteri-Uribarren, Ainitze and Martin, Ana and Casanova, Jorge},
  journal = {Phys. Rev. Res.},
  volume = {7},
  issue = {4},
  pages = {L042016},
  numpages = {6},
  year = {2025},
  month = {Oct},
  publisher = {American Physical Society},
  doi = {10.1103/jksj-736v},
  url = {https://link.aps.org/doi/10.1103/jksj-736v}
}


\newpage







\setcounter{section}{0}
\setcounter{figure}{0}
\setcounter{table}{0}
\setcounter{equation}{0}
\onecolumngrid

\begin{center}
    {\fontsize{18}{60}\bfseries\selectfont Supplementary Information} ~\\[0.5cm]
\end{center}

\renewcommand{\thesection}{S\Roman{section}}
\renewcommand{\theequation}{S.\arabic{equation}}
\renewcommand{\thefigure}{S\arabic{figure}}
\renewcommand{\thetable}{S.\arabic{table}}

\restoreTOC
\tableofcontents

\section{Hamiltonian of the NV center}

The nitrogen-vacancy (NV) center is a spin defect in diamond which comprises a substitutional nitrogen and a vacancy defect. In its negative charge state, this defect is an electron spin-1 system described by the following Hamiltonian:
\begin{equation}\label{SEq. NV Hamiltonian}
    \mathcal{H}_{NV} = \underbrace{DS_z^2 - \gamma_e B_0 S_z}_{\mathcal{H}_0} + \underbrace{A_\parallel S_z I_z + A_\perp (S_x I_x + S_y I_y)}_{\mathcal{H}_{hf}} + \underbrace{Q I_z^2}_{\mathcal{H}_Q} \underbrace{- \gamma_n B_0 I_z}_{\mathcal{H}_{Z}^{(n)}}.
\end{equation}
Here, $D/(2\pi) = 2.88$ GHz is the zero-field splitting, $B_0$ is an external magnetic field aligned the NV quantization axis and $\gamma_e = -28.024 \mathrm{\ GHz/T}$ is the electron gyromagnetic ratio. The term $\mathcal{H}_{hf}$ describes the hyperfine interaction between the electron spin-1 $\mathbf{S}$ and the nitrogen nuclear spin $\mathbf{I}$, where $A_\parallel / (2\pi) = -2.2$ MHz for \ce{^14N} and $A_{\parallel} / (2\pi) = -3.1$ MHz for \ce{^15N}, and $A_\perp$ can be neglected due to the large zero-field splitting of the center. For a \ce{^15N} nucleus with spin $I=1/2$, the quadrupolar interaction $\mathcal{H}_Q$ dissapears, while $Q/(2\pi) = -4.945$ MHz for a \ce{^14N} nucleus with spin $I=1$. The last term in Eq. (\ref{SEq. NV Hamiltonian}) is the nuclear Zeeman interation due to the external field $B_0$, where $\gamma_{\ce{^14N}} / (2\pi) = 3.077 \mathrm{\ MHz/T}$ and $\gamma_{\ce{^15N}} / (2\pi) = -4.316 \mathrm{\ MHz/T}$. Throughout the text, we consider a \ce{^14N} nuclear host for the NV center.

A microwave (MW) driving field can be applied to control the electron spin ($e$) of the NV center, 
\begin{equation}\label{SEq. MW Control}
    \mathcal{H}_{\mathrm{MW}} (t) = \sqrt{2} \Omega_{\mathrm{MW}} \cos(\omega_{\mathrm{MW}} t + \phi_{\mathrm{MW}}) S_x,
\end{equation}
where $\omega_{\mathrm{MW}}$ ($\Omega_{\mathrm{MW}}$) is the (Rabi) frequency of the driving and $\phi_{\mathrm{MW}}$ its initial phase. Tuning the driving frequency to match the lower energy transition $|0\rangle_e \to |\bar{1}\rangle_e$ induces population transfer between these levels, such that the NV center can be effectively treated as a spin-1/2 system. Note that, in this scenario, $S_z = (\sigma_z - \mathbb{1}) / 2$, with $\sigma_z$ the third Pauli matrix (in the basis $\{|0\rangle, |\bar{1}\rangle\}$). Moving to the rotating frame with respect to $\mathcal{H}_0$ and invoking the rotating wave approximation, the MW control Hamiltonian (\ref{SEq. MW Control}) transforms into
\begin{equation}\label{SEq. MW Pulse}
    \mathcal{H}_{\mathrm{MW}}^I = \Omega_{\mathrm{MW}} S_{\phi_{\mathrm{MW}}},
\end{equation}
where $S_{\phi_{\mathrm{MW}}} = \sigma_{\phi_{\mathrm{MW}}}/2$ and $\sigma_{\phi_{\mathrm{MW}}} = \cos\phi_{\mathrm{MW}} \sigma_x + \sin\phi_{\mathrm{MW}} \sigma_y$, with $\sigma_x$ and $\sigma_y$ the first and second Pauli matrices. The propagator associated to the pulse Hamiltonian (\ref{SEq. MW Pulse}), $U_{\phi_{\mathrm{MW}}} = e^{-i\mathcal{H}_{\mathrm{MW}}^I t} = e^{-i \frac{\Omega_{\mathrm{MW}} t}{2} \sigma_{\phi_{\mathrm{MW}}}}$, describes a rotation of an angle $\theta_{\mathrm{MW}} = \Omega_{\mathrm{MW}}t/2$ around the axis set by $\phi_{\mathrm{MW}}$ of the NV state in the Bloch state. 

Similarly, a radio-frequency (RF) driving field may be used to control the nuclear spin ($a$) of the center:
\begin{equation}\label{SEq. RF Control}
    \mathcal{H}_{\mathrm{RF}} (t) = \Omega_{\mathrm{RF}} \cos(\omega_{\mathrm{RF}}t + \phi_{\mathrm{RF}}) I_x,
\end{equation}
where $\omega_{\mathrm{RF}}$ ($\Omega_{\mathrm{RF}}$) is the (Rabi) frequency of the driving and $\phi_{\mathrm{RF}}$ its initial phase. Tuning $\omega_{\mathrm{RF}}$ to match the transition $|0\rangle_a \to |1\rangle_a$, Hamiltonian (\ref{SEq. RF Control}) becomes
\begin{equation}\label{SEq. RF Pulse}
    \mathcal{H}_{\mathrm{RF}} = \Omega_{\mathrm{RF}} I_{\phi_{\mathrm{RF}}},
\end{equation}
with $I_{\phi_{\mathrm{RF}}} = \sigma_{\phi_{\mathrm{RF}}}/2$.

Then, in the following we will work in the two qubit basis, $\{|0\rangle_e |0\rangle_a, |0\rangle_e |1\rangle_a, |\bar{1}\rangle_e |0\rangle_a, |\bar{1}\rangle_e |1\rangle_a\}$, and initialize the system in the $|0\rangle_e |0\rangle_a$ state:
\begin{equation}
    \rho_0 = |0\rangle \langle 0|_e \otimes |0\rangle \langle 0|_a = \left(\begin{array}{cccc}
        1 & 0 & 0 & 0 \\
        0 & 0 & 0 & 0 \\
        0 & 0 & 0 & 0 \\
        0 & 0 & 0 & 0 \\
    \end{array}\right).
\end{equation}

\section{Detailed theory of the gradiometry protocol}\label{SM: Gradiometry}

\subsection{Standard Gradiometry Sequence}

\begin{figure}[b]
    \centering
    \includegraphics[width = 0.6\textwidth]{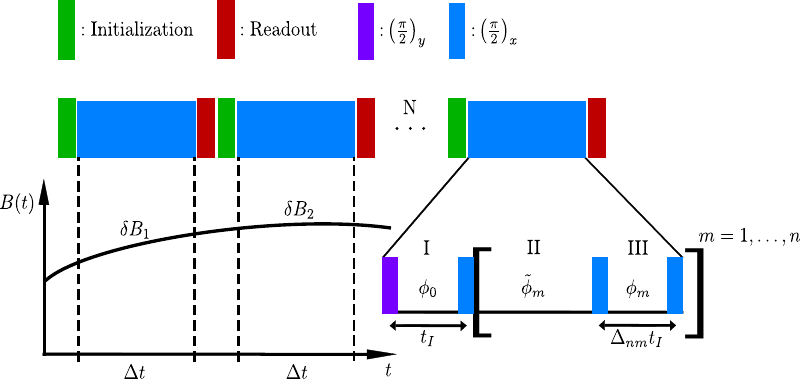}
    \caption{Gradiometric sensing scheme. Our protocol comprises an initialization and a readout step, and a measurement sequence (in green, red and blue, respectively), sequentially repeated $N$ times. Our sequence allows to measure the magnetic field difference between two points separated by a time $\Delta t$. Considering that the magnetic field $B(t)$ changes slowly within a single sequence, each measurement will be proportional to the field difference within the sequence, $\delta B_i = B(t_{i} + \Delta t) - B(t_{i})$, such that, after $N$ repetitions, the full gradient of the field can be recovered. Here, $t_i$ sets the start of $i$-th measurement. Each of these sequences consists of two interrogation stages (I and III) wherein the value of the field remains unchanged and a signal evolution stage (II) wherein the field changes its value. Repetition of stages II and III within each sequence allows access to $n$-th order derivative of the field. Here, the duration of stage III in the $m$-th block (relative to stage I) is given by the combinatorial number $\Delta_{nm} = \left(\begin{array}{c}
    n \\
    m \\
\end{array}\right)$.}
    \label{SF: Sequence}
\end{figure}

Our gradiometry sequence comprises three main stages: two interrogation stages and a signal evolution stage (see Fig. \ref{SF: Sequence}). During the first interrogation stage (I), a $(\pi/2)_y$-pulse takes the sensor state to a superposition. Then, the sensor accumulates some phase $\phi_0$ due to a quasi-static magnetic field $B(t)$. Next, a $(\pi/2)_x$-pulse encodes the coherence information into qubit populations, such that the state of the sensor reads:
\begin{equation}
    \rho_0 = \left(\begin{array}{cc} 
        1 & 0 \\
        0 & 0 \\
    \end{array}\right) \to \rho_I = \frac{1}{2}\left(\begin{array}{cc}
        1 + \sin\phi_0 & \cos\phi_0 \\
        \cos\phi_0 & 1 - \sin\phi_0 \\
    \end{array}\right).
\end{equation}
A stage of signal evolution (II) follows: the sensor accumulates a phase $\tilde{\phi}_1$ due to the time varying field $B(t)$, and its state becomes:
\begin{equation}
    \rho_{II} = \frac{1}{2}\left(\begin{array}{cc}
        1 + \sin\phi_0 & e^{-i \tilde{\phi}_1} \cos\phi_0 \\
        e^{i \tilde{\phi}_1} \cos\phi_0 & 1 - \sin\phi_0
    \end{array}\right).
\end{equation}
During the second interrogation stage (III), the field becomes quasi-static again, although its value might different than that of stage I due to the intermediate signal evolution stage. The sensor accumulates a phase $\phi_1$ during this interrogation stage and, at the end of the sequence, its state is written as 
\begin{equation}
    \rho_{III} = \frac{1}{2} \left(\begin{array}{cc}
        1 + \sin\phi_1 \cos\tilde{\phi}_1 \cos\phi_0 - \cos\phi_1 \sin\phi_0 & i \sin\tilde{\phi}_1 \cos\phi_0 + \cos\phi_1 \cos\tilde{\phi}_1 \cos\phi_0 + \sin\phi_1 \sin\phi_0 \\
        -i \sin\tilde{\phi}_1 \cos\phi_0 + \cos\phi_1 \cos\tilde{\phi}_1 \cos\phi_0 + \sin\phi_1 \sin\phi_0 & 1 - \sin\phi_1 \cos\tilde{\phi}_1 \cos\phi_0 + \cos\phi_1 \sin\phi_0
    \end{array}\right).
\end{equation}
Upon readout of the sensor state, we find
\begin{equation}
    \langle \sigma_z \rangle = \mathrm{Tr}(\rho_{III} \sigma_z) = \sin\phi_1 \cos\tilde{\phi}_1 \cos\phi_0 - \cos\phi_1 \sin\phi_0,
\end{equation}
which for small $\tilde{\phi}_1$ can be approximated as
\begin{equation}\label{SEq. Derivative}
    \langle \sigma_z \rangle \approx \sin(\phi_1 - \phi_0).
\end{equation}
For DC magnetometry, the target magnetic field $B(t)$ will typically comprise two parts: a static background field $B$ and a slowly varying signal $b(t)$, which we want to detect, \textit{i.e.}, $B(t) = B + b(t)$. Standard DC sensing techniques, such as the Ramsey sequence (stage I in Fig. \ref{SF: Sequence}) or, in the case of color centers, optically detected magnetic resonance (ODMR), rely on total intensity measurements, such that magnetic shielding from the environment is often required to suppress static fields or magnetic drifts. Our protocol intrinsically suppresses the effect of these, as long as the phase $\tilde{\phi}_1$ is small. Then, for a given measurement starting at time $t_i$, $\phi_1 - \phi_0 \propto (B + b_{III}^i) - (B + b_I^i) = b_{III}^i - b_I^i$, where $b_I^i = b(t_i)$ and $b_{III}^i = b(t_i + \Delta t)$, with $\Delta t = t_{II}$. Then, considering $N$ consecutive differential measurements, the discretized temporal gradient of the field can be reconstructed,
\begin{equation}
    \left.\frac{d B(t)}{dt}\right|_{t_i} = \left.\frac{db(t)}{dt}\right|_{t_i} \approx \frac{\delta b_i}{\Delta t},
\end{equation}
where $\delta b_i = b_{III}^i - b_I^i$ is the magnetic field difference obtained in the $i$-th measurement stage (see Fig. \ref{SF: Sequence}).

\subsection{Generalized sequence}

Our gradiometric sequence can be generalized to measure higher order derivatives. By repeating stages II and III a total of $n$ times, the $n$-th order derivative can be measured (see Fig. \ref{SF: Sequence}). We may write the sensor state at the end of the $(n-1)$-th block as 
\begin{equation}
    \rho_{n-1} = \frac{\mathbb{1}}{2} + \frac{1}{2}\mathbf{v}_{n-1}\cdot\bm{\sigma},
\end{equation} 
where $\mathbf{v}_{n-1} = (a_{n-1}, b_{n-1}, c_{n-1})$ is the Bloch vector and $\bm{\sigma} = (\sigma_x, \sigma_y, \sigma_z)$. It is useful to consider how the Pauli matrices transform under rotations around different axes. For a $(\pi/2)$ rotation around the $x$ axis, $\sigma_x \to \sigma_x$, $\sigma_y \to \sigma_z$ and $\sigma_z \to -\sigma_y$; and around the $y$ axis, $\sigma_x \to -\sigma_z$, $\sigma_y \to \sigma_y$ and $\sigma_z \to \sigma_x$. For a $\varphi$ rotation around the $z$ axis, $\sigma_x \to \cos\varphi \sigma_x + \sin\varphi \sigma_y$, $\sigma_y \to \cos\varphi \sigma_y - \sin\varphi \sigma_x$ and $\sigma_z \to \sigma_z$.

Then, during stage II in the $n$-th block, the sensor accumulates a phase $\tilde{\phi}_n$ such that its state reads
\begin{equation}
    \rho_{n-1}^{(II)} = \frac{\mathbb{1}}{2} + \frac{1}{2}\left[ a_{n-1} \left(\cos\tilde{\phi}_n \sigma_x + \sin\tilde{\phi}_n \sigma_y\right) + b_{n-1} \left(\cos\tilde{\phi}_n \sigma_y - \sin\tilde{\phi}_n \sigma_x\right) + c_{n-1} \sigma_z\right].
\end{equation}
Next, the sensor undergoes the $n$-th stage III, during which it accumulates a phase $\phi_n$, and the state before measurement reads
\begin{align}
    \rho_{n-1}^{(III)} = \frac{\mathbb{1}}{2} + \frac{1}{2} & \left[\left(\cos\tilde{\phi}_n \cos\phi_n a_{n-1} - \sin\tilde{\phi}_n \cos\phi_n b_{n-1} + \sin\phi_n c_{n-1}\right) \sigma_x \right. \notag\\
    + & \left(-\sin\tilde{\phi}_n a_{n-1} - \cos\tilde{\phi}_n b_{n-1} \right) \sigma_y \notag\\
    + & \left. \left(\cos\tilde{\phi}_n \sin\phi_n a_{n-1} - \sin\tilde{\phi}_n \sin\phi_n b_{n-1} - \cos\phi_n c_{n-1}\right) \sigma_z \right] = \rho_n,
\end{align}
from where we extract recursive relations for the components of the Bloch vector:
\begin{equation}
    \begin{array}{ccl}
        a_n & = & \cos\tilde{\phi}_n \cos\phi_n a_{n-1} - \sin\tilde{\phi}_n \cos\phi_n b_{n-1} + \sin\phi_n c_{n-1}, \\
        b_n & = & -\sin\tilde{\phi}_n a_{n-1} - \cos\tilde{\phi}_n b_{n-1}, \\
        c_n & = & \cos\tilde{\phi}_n \sin\phi_n a_{n-1} - \sin\tilde{\phi}_n \sin\phi_n b_{n-1} - \cos\phi_n c_{n-1},
    \end{array}
\end{equation}
with $a_0 = \sin\phi_0$, $b_n = 0$, $c_n = \cos\phi_0$. Trivially, we find that $\langle \sigma_z \rangle = \mathrm{Tr}(\rho_n \sigma_z) = c_n$, which for small $\tilde{\phi}_n$ can be written as 
\begin{equation}\label{SEq. Gradiometry}
    \langle \sigma_z \rangle = c_n \approx (-1)^n \sin\left[\sum_{m=0}^n (-1)^m \phi_m\right].
\end{equation}
To obtain expression (\ref{SEq. Gradiometry}), we first consider the coefficients $a_n$, $b_n$ and $c_n$ under the approximation $\tilde{\phi}_n \approx 0$:
\begin{align}
    a_n & \approx \cos\phi_n a_{n-1} + \sin\phi_n c_{n-1}, \notag \\
    b_n & \approx 0, \\
    c_n & \approx \sin\phi_n a_{n-1} - \cos\phi_n c_{n-1} \notag.
\end{align}
Then, $c_n$ can be expanded as
\begin{align}
    c_n & = \sin\phi_n a_{n-1} - \cos\phi_n c_{n-1} \notag\\
    & = \sin(\phi_n - \phi_{n-1}) a_{n-2} + \cos(\phi_n - \phi_{n-1}) c_{n-2} \notag\\
    & = \sin(\phi_n - \phi_{n-1} + \phi_{n-2}) a_{n-3} - \cos(\phi_n - \phi_{n-1} + \phi_{n-2}) c_{n-3} \\
    & = \dotsc \notag
\end{align}
where angle sum and difference trigonometric identities have been used. Easily, one realizes that, for odd $n$, we find
\begin{equation}
    c_n = \sin[\phi_n - \phi_{n-1} + \dotsb + \phi_1] a_0 - \cos[\phi_n - \phi_{n-1} + \dotsb + \phi_1] c_0 = \sin[\phi_n - \phi_{n-1} + \dotsb + \phi_1 - \phi_0] = (-1) \sin\left[\sum_{m=0}^n (-1)^m \phi_m \right],
\end{equation}
while for even $n$
\begin{equation}
    c_n = \sin[\phi_n - \phi_{n-1} + \dotsb - \phi_1] a_0 + \cos[\phi_n - \phi_{n-1} + \dotsb - \phi_1] c_0 = \sin[\phi_n - \phi_{n-1} + \dotsb - \phi_1 + \phi_0] = (+1) \sin\left[\sum_{m=0}^n (-1)^m \phi_m \right],
\end{equation}
which can be combined into Eq. (\ref{SEq. Gradiometry}).

Now, note that the duration of the $m$-th stage III in the $n$-th order protocol is chosen to be $t_{III, m}^{(n)} = \Delta_{nm} t_I$, where
\begin{equation}
    \Delta_{nm} = \left(\begin{array}{c}
n \\
m \\
\end{array}\right)
\end{equation}
is the combinatorial number. Then, Eq. (\ref{SEq. Gradiometry}) can be approximated as 
\begin{equation}
    \langle \sigma_z \rangle^{(n)} \approx \sum_{m=0}^n (-1)^{n+m} \phi_m \propto (-1)^n b_I + \sum_{m=1}^n (-1)^{n+m} \Delta_{nm} b_{III, m} = \delta b,
\end{equation}
where $b_I$ and $b_{III, m}$ are the values of the signal $b(t)$ during stage I and the $m$-th stage III, respectively. Again, we see that the background field $B$ is suppressed. As before, by performing $N$ of such measurements, the discretized $n$-th order time derivative of the total field $B(t) = B + b(t)$ can be obtained:
\begin{equation}
    \left.\frac{d^{(n)} B (t)}{dt^n}\right|_{t_i} \approx \frac{\delta b_i}{\Delta t^n},    
\end{equation}
with $\Delta t$ the duration of stage II and the superscript $i$ referring to the $i$-th measurement. As an example, consider the second order protocol; here, $\Delta_{21} = 2$ and $\Delta_{22} = 1$, so that we find
\begin{equation}
    \langle \sigma_z \rangle^{(2)} \propto b_I - 2b_{III, 1} + b_{III, 2} = \delta b,
\end{equation}
and thus
\begin{equation}
    \left.\frac{d^{(2)}B(t)}{dt^2}\right|_{t_i} = \left.\frac{d^{(2)}b(t)}{dt^2}\right|_{t_i} \approx \frac{b_I^i - 2b_{III, 1}^i + b_{III, 2}^i}{t_{II}^2},
\end{equation}
which is the discretized second order derivative of the field $B(t)$.

\subsection{Introducing a nuclear memory}

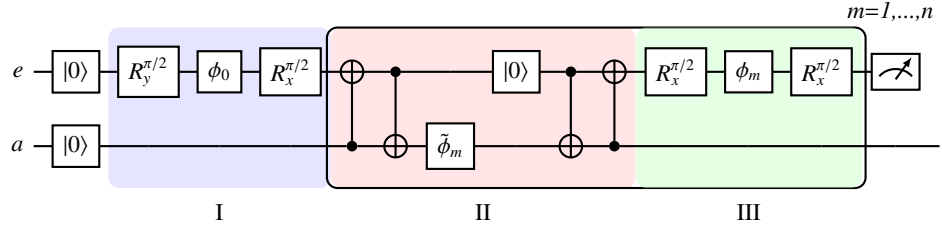
\begin{figure}[t]
    \centering
    \begin{quantikz}[row sep=0.35cm, column sep=0.25cm]
        \lstick{\textit{e}} 
        & \gate{\ket{0}}
        & \gate{R_y^{\pi/2}}\gategroup[wires=2,steps=3,style={fill=blue!20, inner xsep=0.5pt, draw=none, fill opacity=0.5, rounded corners}, background, label style={label position=below,anchor=center,yshift=-0.5cm}]{\text{I}}
        & \gate{\phi_0}
        & \gate{R_x^{\pi/2}}
        & \targ{}\gategroup[wires=2,steps=6,style={fill=red!20, inner xsep=0.5pt, draw=none, fill opacity=0.5, rounded corners}, background, label style={label position=below,anchor=center,yshift=-0.5cm}]{\text{II}}\gategroup[wires=2,steps=9,style={inner xsep=2pt, draw=black, fill opacity=0.5, rounded corners}, background, label style={label position=above right,anchor=east,yshift=0cm,xshift=3.5cm}]{\text{\textit{m=1,...,n}}}
        & \ctrl{1}
        & \qw
        & \gate{\ket{0}}
        & \ctrl{1}
        & \targ{}
        & \gate{R_{x}^{\pi/2}}\gategroup[wires=2,steps=3,style={fill=green!20, inner xsep=0.5pt, draw=none, fill opacity=0.5, rounded corners}, background, label style={label position=below,anchor=center,yshift=-0.5cm}]{\text{III}}
        & \gate{\phi_m}
        & \gate{R_{x}^{\pi/2}}
        & \meter{}
        \\
        \lstick{\textit{a}}
        & \gate{\ket{0}}
        & \qw
        & \qw
        & \qw
        & \ctrl{-1}
        & \targ{}
        & \gate{\tilde{\phi}_m}
        & \qw
        & \targ{}
        & \ctrl{-1}
        & \qw
        & \qw
        & \qw
        & \qw
        & \qw
    \end{quantikz}
    \label{SF: Memory}
    \caption{General sequence for measuring the $n$-th order derivative of the field including an ancillary qubit serving as a nuclear memory. SWAP gates are used during stage II for encoding/recovering the sensor ($e$) state in/from the memory ($a$) such that the protocol is limited by the nuclear dephasing time $T_2^a$. Stages II and III are repeated $n$ times, with the duration of the $m$-th repetition of stage III given by $t_{III,m}^{(n)} = \Delta_{nm} t_I$, where $\Delta_{nm} = \left(\begin{array}{c}
        n \\
        m\end{array}\right)$.}
\end{figure}

The gradiometry protocol proposed works provided the phases $\tilde{\phi}_n$ are small, as the approximation in Eqs. (\ref{SEq. Derivative}) and (\ref{SEq. Gradiometry}) does not hold when the duration of the $n$-th signal evolution stage exceeds the sensor dephasing time $T_2^*$. Luckily, because we want this phase to have no effect on the final result, we can apply refocusing pulses such that this limitation is extended to the coherence time $T_2$. However, this may not suffice if the target signal changes in a timescale larger than $T_2$. In these cases, we may transfer the phase accumulation process of stage II to a nuclear ancillary qubit or nuclear memory through SWAP gates (see Fig. \ref{SF: Memory}). These can easily be constructed by the consecutive application of CNOT gates.

A weak MW pulse ($\Omega_{\mathrm{MW}} \ll A_\parallel$) resonant with the $|0\rangle_e |1\rangle_a \to |\bar{1}\rangle_e |1\rangle_a$ transition results in a controlled-rotation operation, with the control being the memory and the target the sensor, \textit{i.e.}, a $\mathrm{C}_a\mathrm{ROT}_e$ gate, which is described by
\begin{equation}\label{SEq. CnROTe}
    U_{\mathrm{C}_a\mathrm{ROT}_e}^{\phi_{\mathrm{MW}}}(\theta_{\mathrm{MW}}) = \mathbb{1} \otimes |0\rangle \langle 0|_a + e^{-i \frac{\theta_{\mathrm{MW}}}{2} \sigma_{\phi_{\mathrm{MW}}}} \otimes |1\rangle\langle1|_a,
\end{equation}
where the angle of rotation is $\theta_{\mathrm{MW}} = \Omega_{\mathrm{MW}} t_p$, $t_p$ being the duration of the pulse, and $\phi_{\mathrm{MW}}$ is the pulse axis. For the particular case where $\theta_{\mathrm{MW}} = \pi$, the propagator (\ref{SEq. CnROTe}) described a $\mathrm{C}_a\mathrm{NOT}_e$-gate:
\begin{equation}\label{SEq. CnNOTe}
    U_{\mathrm{C}_a\mathrm{NOT}_e} = \left(\begin{array}{cccc}
        1 & 0 & 0 & 0 \\
        0 & 0 & 0 & -i \\
        0 & 0 & 1 & 0 \\
        0 & -i & 0 & 0 \\
    \end{array}\right),
\end{equation}
where we have taken $\phi_{\mathrm{MW}} = 0$ for simplicity.

Similarly, a weak RF pulse ($\Omega_{\mathrm{RF}} \ll A_\parallel$) resonant with the $|\bar{1}\rangle_e|0\rangle_a \to |\bar{1}\rangle_e|1\rangle_a$ transition results in a controlled-rotation operation, with the sensor as the control and the memory as the target, \textit{i.e.}, a $\mathrm{C}_e\mathrm{ROT}_a$ gate, described by 
\begin{equation}\label{SEq. CeROTn}
    U_{\mathrm{C}_e\mathrm{ROT}_a}^{\phi_{\mathrm{RF}}}(\theta_{\mathrm{RF}}) = |0\rangle\langle0|_e \otimes \mathbb{1} + |\bar{1}\rangle\langle\bar{1}|_e \otimes e^{-i \frac{\theta_{\mathrm{RF}}}{2} \sigma_{\phi_{\mathrm{RF}}}},
\end{equation}
where $\theta_{\mathrm{RF}}$ and $\phi_{\mathrm{RF}}$ are defined analogously to the previous case. Again, for the particular case $\theta_{\mathrm{RF}} = \pi$, Eq. (\ref{SEq. CeROTn}) describes a $\mathrm{C}_e\mathrm{NOT}_a$-gate:
\begin{equation}\label{SEq. CeNOTn}
    U_{\mathrm{C}_e\mathrm{NOT}_a} = \left(\begin{array}{cccc}
        1 & 0 & 0 & 0 \\
        0 & 1 & 0 & 0 \\
        0 & 0 & 0 & -i \\
        0 & 0 & -i & 0 \\
    \end{array}\right),
\end{equation}
where, for simplicity, $\phi_{\mathrm{RF}} = 0$.

The alternating concatenation of three $\mathrm{C}_e\mathrm{NOT}_a/\mathrm{C}_a\mathrm{NOT}_e$ gates results in a SWAP gate:
\begin{equation}
    U_{\mathrm{SWAP}} = U_{\mathrm{C}_e\mathrm{NOT}_a} U_{\mathrm{C}_a\mathrm{NOT}_e} U_{\mathrm{C}_e\mathrm{NOT}_a} = U_{\mathrm{C}_a\mathrm{NOT}_e} U_{\mathrm{C}_e\mathrm{NOT}_a} U_{\mathrm{C}_a\mathrm{NOT}_e} = \left(\begin{array}{cccc}
        1 & 0 & 0 & 0 \\
        0 & 0 & -1 & 0 \\
        0 & -1 & 0 & 0 \\
        0 & 0 & 0 & -1
    \end{array}\right).
\end{equation}
Consider the initial two qubit state $\rho_0 = \rho_e \otimes \rho_a$, where $\rho_e$ takes an arbitrary form while $\rho_a = |0\rangle\langle 0|_a$:
\begin{equation}
    \rho_0 = \left(\begin{array}{cc}
        \rho_{00}^e & \rho_{01}^e \\
        \rho_{10}^e & \rho_{11}^e 
    \end{array} \right) \otimes |0\rangle\langle 0|_a = \rho_{00}^e |00\rangle\langle 00| + \rho_{01}^e |00\rangle\langle\bar{1}0| + \rho_{10}^e |\bar{1}0\rangle\langle 00| + \rho_{11}^e |\bar{1}0\rangle\langle\bar{1}0|,
\end{equation}
where, for the sake of clarity, we write, for instance, $|00\rangle \langle 00| = |0\rangle\langle0|_e \otimes |0\rangle\langle0|_a$. Then, under the application of a SWAP gate $\rho_0$ is transformed as follows:
\begin{equation}\label{SEq. SwappedState}
    \rho_1 = U_{\mathrm{SWAP}} \rho_0 U_{\mathrm{SWAP}}^\dagger = \rho_{00}^e |00\rangle\langle 00| - \rho_{01}^e |00\rangle\langle 01| - \rho_{10}^e |01\rangle\langle 00| + \rho_{11}^e |01\rangle\langle 01|,
\end{equation}
which may be rewritten as 
\begin{equation}
    \rho_0 = |0\rangle\langle 0|_e \otimes \left(\begin{array}{cc}
        \rho_{00}^e & -\rho_{01}^e \\
        -\rho_{10}^e & \rho_{11}^e 
    \end{array} \right).
\end{equation}
Notice that the coherences acquire an additional negative sign which, however, is not relevant for our purposes as after applying a second SWAP gate, we recover the initial state $\rho_0$. 

Alternatively, in the case we are considering, where the memory is initialized in the $|0\rangle_n$ state, we can effectively generate two-gate SWAP operations. For encoding into the memory, we find that
\begin{equation}
    U_{\mathrm{C}_a\mathrm{NOT}_e} U_{\mathrm{C}_e\mathrm{NOT}_a} = \left(\begin{array}{cccc}
        1 & 0 & 0 & 0 \\
        0 & 0 & -1 & 0 \\
        0 & 0 & 0 & -i \\
        0 & -i & 0 & 0 \\
    \end{array}\right)
\end{equation}
produces the same state as in Eq. (\ref{SEq. SwappedState}), and then
\begin{equation}
    U_{\mathrm{C}_e\mathrm{NOT}_a} U_{\mathrm{C}_a\mathrm{NOT}_e} = \left(\begin{array}{cccc}
        1 & 0 & 0 & 0 \\
        0 & 0 & 0 & -i \\
        0 & -1 & 0 & 0 \\
        0 & 0 & -i & 0 \\
    \end{array}\right)
\end{equation}
recovers the initial state.

\subsection{Modified sequence for $T_1$-limited gradiometry}

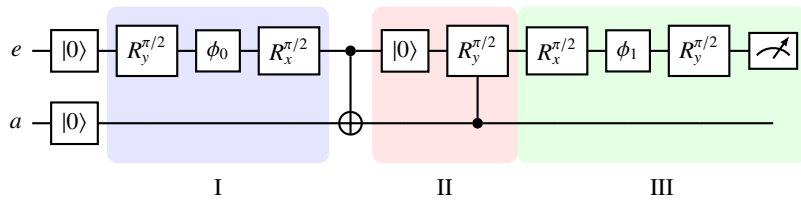
\begin{figure}[b]
    \centering
    \begin{quantikz}[row sep=0.35cm, column sep=0.25cm]
        \lstick{\textit{e}} 
        & \gate{\ket{0}}
        & \gate{R_y^{\pi/2}}\gategroup[wires=2,steps=3,style={fill=blue!20, inner xsep=0.5pt, draw=none, fill opacity=0.5, rounded corners}, background, label style={label position=below,anchor=center,yshift=-0.5cm}]{\text{I}}
        & \gate{\phi_0}
        & \gate{R_x^{\pi/2}}
        & \ctrl{1}
        & \gate{\ket{0}}\gategroup[wires=2,steps=2,style={fill=red!20, inner xsep=0.5pt, draw=none, fill opacity=0.5, rounded corners}, background, label style={label position=below,anchor=center,yshift=-0.5cm}]{\text{II}}
        & \gate{R_y^{\pi/2}}
        & \gate{R_{x}^{\pi/2}}\gategroup[wires=2,steps=4,style={fill=green!20, inner xsep=0.5pt, draw=none, fill opacity=0.5, rounded corners}, background, label style={label position=below,anchor=center,yshift=-0.5cm}]{\text{III}}
        & \gate{\phi_1}
        & \gate{R_{y}^{\pi/2}}
        & \meter{}
        \\
        \lstick{\textit{a}}
        & \gate{\ket{0}}
        & \qw
        & \qw
        & \qw
        & \targ{}
        & \qw
        & \ctrl{-1}
        & \qw
        & \qw
        & \qw
        & \qw
    \end{quantikz}
    \label{SF: 1c}
    \caption{Modified protocol for $T_1^a$-limited magnetometry. Only the population information is encoded in the nuclear memory during stage II.} 
\end{figure}

The protocol in Fig. \ref{SF: Memory} can be modified such that stage II is limited by the nuclear relaxation time $T_1^a$, which is typically several orders of magnitude longer than the dephasing time $T_2^a$. Here, the SWAP gate at the beggining of stage II is replaced by a $\mathrm{C}_e\mathrm{NOT}_a$ gate, hence only encoding the population information in the nuclear memory. The state of the system reads
\begin{equation}
    \rho = \frac{1}{2}\left(\begin{array}{cc}
        1 + \sin\phi_0 & 0 \\
        0 & 1 - \sin\phi_0 
    \end{array}\right) \otimes \frac{1}{2}\left(\begin{array}{cc}
        1 + \sin\phi_0 & 0 \\
        0 & 1 - \sin\phi_0 
    \end{array}\right).
\end{equation}
Next, the sensor is reinitialized into the $|0\rangle_e$ state and a $\mathrm{C}_a\mathrm{ROT}_e^y(\pi/2)$ gate is applied to recover the state from the memory, such that
\begin{equation}
    \rho = \frac{1}{4}\left(\begin{array}{cc}
        3 + \sin\phi_0 & 1 - \sin\phi_0 \\
        1 - \sin\phi_0 & 1 - \sin\phi_0 
    \end{array}\right) \otimes \frac{1}{2}\left(\begin{array}{cc}
        1 + \sin\phi_0 & 0 \\
        0 & 1 - \sin\phi_0 
    \end{array}\right).
\end{equation}
Finally, after stage III, the expected value of $\sigma_z$ reads
\begin{equation}
    \langle \sigma_z \rangle = \mathrm{Tr}(\rho_{III} \sigma_z) = -\frac{1}{2} ( \cos\phi_1 + \sin\phi_1 - \sin\phi_0 \cos\phi_1 + \sin\phi_0 \sin\phi_1),
\end{equation}
which, assuming small $\phi_i$ can be approximated as
\begin{equation}
    \langle \sigma_z \rangle \approx -\frac{1}{2}\left[1 + (\phi_1 - \phi_0)\right].
\end{equation}
Because only the population information is encoded into the nuclear memory, no phase is accumulated during stage II, allowing longer signal evolution times, limited only by the nuclear relaxation time.

\section{Detailed theory of oscillating NV}\label{SM: DCtoAC}

Inducing an effective AC signal with a high frequency such as $\nu/(2\pi) = 1$ MHz is not achievable via mechanical rotation. Thus, we propose inducing small oscillations of the sensor, which allows much larger frequencies. Consider a static field the laboratory frame, $\bm{B} = (B_x, B_y, B_z)$. We can write the field in the oscillating frame, \textit{i.e.}, the frame of the NV, as 
\begin{equation}
    \mathbf{B}^{\mathrm{NV}}(t) = R_{\hat{k}}(\theta_{NV}(t)) \mathbf{B},
\end{equation}
where $\hat{k} = (k_x, k_y, k_z)$ is a unit vector which defines the axis of oscillation and $\theta_{NV}(t)$ is the angle between the $z$ and $z^{\mathrm{NV}}$ axes. This is given by $\theta_{NV}(t) = \theta_0 + \alpha \sin\nu t$, with $\theta_0$ the initial angle between both axes, and $\alpha$ and $\nu$ the amplitude and frequency of the oscillations. The general rotation matrix is given by 
\begin{equation}
    R_{\hat{k}}(\theta) = \left(\begin{array}{ccc}
        k_x^2 (1 - \cos\theta) + \cos\theta & k_x k_y (1 - \cos\theta) - k_z \sin\theta & k_x k_z(1 - \cos\theta) + k_y \sin\theta \\
        k_x k_y (1 - \cos\theta) + k_z \sin\theta & k_y^2 (1 - \cos\theta) + \cos\theta & k_y k_z (1 - \cos\theta) - k_x \sin\theta \\
        k_x k_z (1 - \cos\theta) - k_y \sin\theta & k_y k_z (1 - \cos\theta) + k_x \sin\theta & k_z^2 (1 - \cos\theta) + \cos\theta \\
    \end{array}\right),
\end{equation}
such that the component of the field that is sensed by the NV center (in the reference frame of the sensor) reads
\begin{align}
    B_z^{\mathrm{NV}} (t) = [k_x k_z (1 - \cos\theta_{NV}(t)) - k_y \sin\theta_{NV}(t)] B_x + [k_y k_z (1 - \cos\theta_{NV}(t)) + k_x \sin\theta_{NV}(t)] B_y + [k_z^2 (1 - \cos\theta_{NV}(t)) + \cos\theta_{NV}(t)] B_z.
\end{align}
Because we assume a small oscillation amplitude $\alpha$, we can approximate
\begin{equation}
    \begin{array}{ccc}
        \sin\theta_{NV}(t) & \approx & \sin\theta_0 + \alpha \cos\theta_0 \sin\nu t, \\
        \cos\theta_{NV}(t) & \approx & \cos\theta_0 - \alpha \sin\theta_0 \sin\nu t, \\
    \end{array}
\end{equation}
and knowing that static terms will have no effect as we will apply dynamical decoupling, we find
\begin{align}
    B_z^{\mathrm{NV}} (t) \approx \alpha \sin\nu t [k_x k_z \sin\theta_0 - k_y \cos\theta_0] B_x + \alpha \sin\nu t [k_y k_z \sin\theta_0 + k_x \cos\theta_0] B_y + \alpha \sin\theta_0 \sin\nu t (k_z^2 - 1) B_z.
\end{align}
Now, by setting $\theta_0 = 0$, 
\begin{equation}
    B_z^{\mathrm{NV}} (t) = \alpha \sin\nu t (k_x B_y - k_y B_x),
\end{equation}
such that $B_z^{\mathrm{NV}} (t) = \alpha \sin\nu t B_y$ for $k_x = 1$ and $k_y = 0$, and $B_z^{\mathrm{NV}} (t) = -\alpha \sin\nu t B_x$ for $k_x = 0$ and $k_y = 1$. Similarly, setting $\theta_0 = \pi/2$,
\begin{equation}
    B_z^{\mathrm{NV}} (t) = \alpha \sin\nu t (k_x k_z B_x + k_y k_z B_y) + \alpha \sin\nu t (k_z^2 - 1) B_z,
\end{equation}
which is zero for $k_{x, y} = 0$ and $k_z = 1$, and $B_z^{\mathrm{NV}} (t) = -\alpha \sin\nu t B_z$ for $\hat{k}$ in the $xy$ plane.

The evolution of the sensor spin state due to the effective AC signal induced by the oscillation is thus described by 
\begin{equation}
    \mathcal{H}(t) = |\gamma_e| B_z^{\mathrm{NV}}(t) S_z = \alpha |\gamma_e| \sin\nu t B_i S_z,
\end{equation}
where $B_i = -B_x, B_y, -B_z$. To capture this signal, we apply a dynamical decoupling sequence such that the evolution operator in the toggling frame reads
\begin{equation}
    U = \exp(-i \mathcal{H}_{\mathrm{eff.}} (2\tau N_p)),
\end{equation}
where $\tau = 1/(4\nu)$ is the interpulse separation, $N_p$ the number of pulses in the sequence and
\begin{equation}
    \mathcal{H}_{\mathrm{eff.}} = \frac{S_z}{2\tau N_p} \int_{0}^{2\tau N_p} \mathcal{F}(t) \mathcal{H}(t) \ dt = \frac{2\alpha |\gamma_e| B_i N_p}{\nu} S_z.
\end{equation}
Here, $\mathcal{F}(t)$ is an odd modulation function imprinted by the dynamical decoupling sequence, which is unity at $t=0$ and changes sign when every time a $\pi$-pulse is applied. Then, the phase accumulated by the sensor is 
\begin{equation}
    \phi = \frac{2\alpha N_p |\gamma_e| B_i}{\nu}.
\end{equation}

\section{Modelling of magnetic anomalies}\label{SM: MagAnomalies}

In order to model magnetic anomalies, we will assume that the measurement occurs in the far-field regime, such that the source of the anomaly, which is considered to be a sphere, can be approximated as a dipole:
\begin{equation}
    \mathbf{B}_{\mathrm{an.}} (\mathbf{r}) = \frac{\mu_0}{4\pi} \left(\frac{3(\mathbf{m}\cdot \mathbf{r}) \mathbf{r}}{r^5} - \frac{\mathbf{m}}{r^3}\right),
\end{equation}
where $\mu_0$ is the magnetic susceptibility of vacuum and $\mathbf{m}$ is the magnetic dipole moment, which is related to the magnetization $\mathbf{M}$ of the source by $\mathbf{M} = \mathbf{m} / V$, with $V$ its volume. This magnetization is due to an external or inducing field $\mathbf{H}_{\mathrm{ext.}}$ (in this case, this is the geomagnetic field) and an internal field $\mathbf{H}_{\mathrm{int.}}$ which originates from the induced magnetization, such that
\begin{equation}
    \mathbf{M} = \chi \mathbf{H} = \chi (\mathbf{H}_{\mathrm{ext.}} + \mathbf{H}_{\mathrm{int.}}),
\end{equation}
with $\chi$ the magnetic susceptibility of the source. In this case where the inducing field is the geomagnetic field, we have that $\mathbf{H}_{\mathrm{ext.}} = \mathbf{B}_{\mathrm{gm}} / \mu_0$. The internal field, on the other hand, is given by the following expression:
\begin{equation}
    \mathbf{H}_{\mathrm{int.}} (\mathbf{r}) = \frac{1}{4\pi} \int_{V'} \mathbf{M}(\mathbf{r'}) \mathbf{\nabla}^2 \frac{1}{|\mathbf{r} - \mathbf{r}'|} \ d^3\mathbf{r}'.
\end{equation}
For a uniformly magnetized sphere, it can be easily shown that
\begin{equation}
    \mathbf{M} = \frac{3\chi}{3 + \chi}\mathbf{H}_{\mathrm{ext.}},
\end{equation}
which in the limit of high susceptibility (\textit{e.g.}, for a ferromagnetic source), $\mathbf{M} \to 3\mathbf{H}_{\mathrm{ext.}}$. Then, the magnetic field emitted by the source, \textit{i.e.}, the magnetic anomaly, can be described by
\begin{equation}\label{SMEq. AnomalyField}
    \mathbf{B}(\mathbf{r}) = \frac{V}{4\pi r^3} \frac{3\chi}{3 + \chi} \left[ 3(\mathbf{H}_{\mathrm{ext.}} \cdot \mathbf{\hat{r}}) \mathbf{\hat{r}} - \mathbf{H}_{\mathrm{ext.}} \right].
\end{equation}
The figure shows the magnetic field emitted by a spherical ferromagnetic source of radius $R = 16$ cm, calculated using Eq. (\ref{SMEq. AnomalyField}) over a square of length $2L = 10$m at a height $h = 1$m. 

\begin{figure*}[ht]
    \subfloat[\label{SF:3a}]{%
    \includegraphics[width=0.313\textwidth]{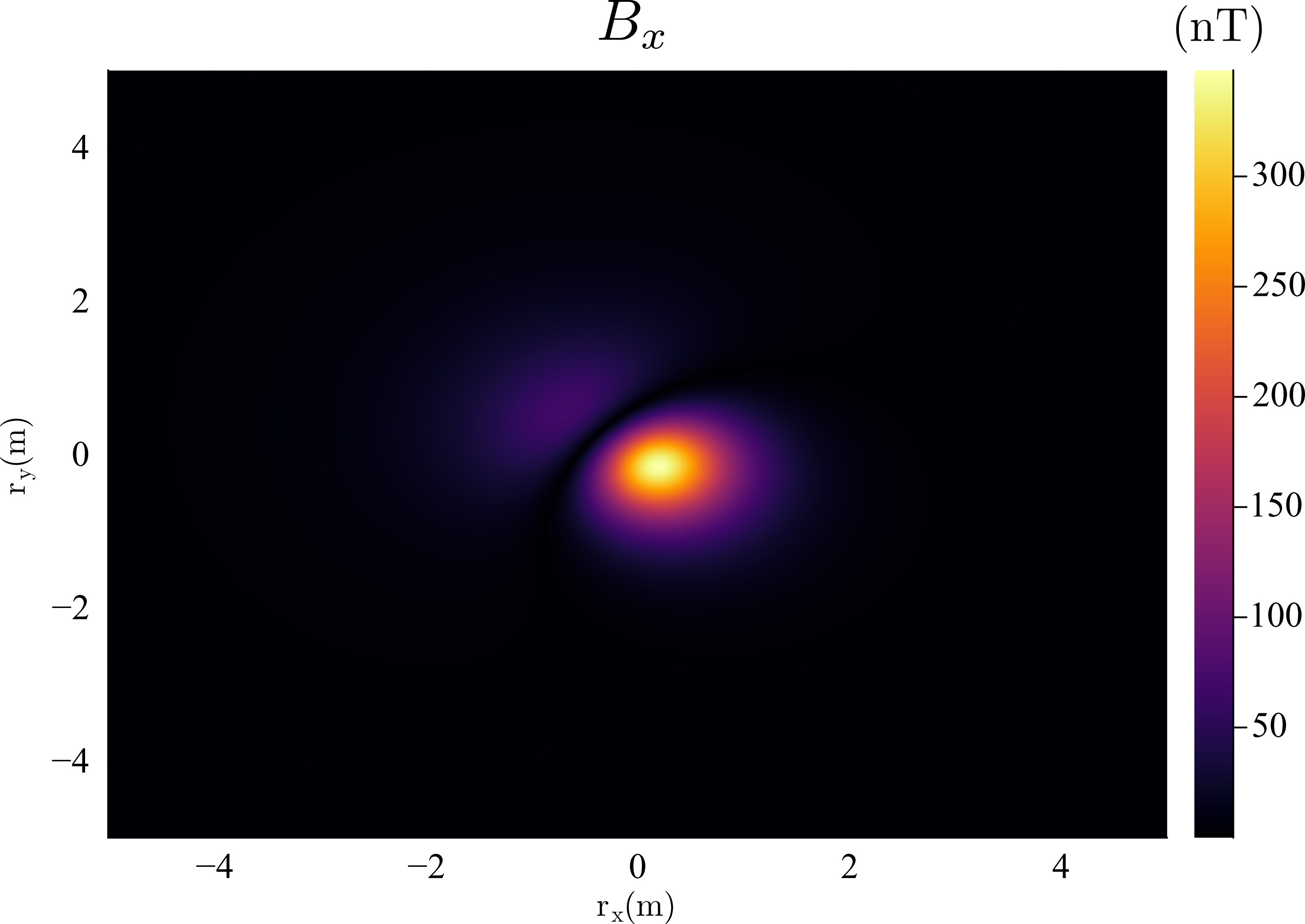}%
    }
    \subfloat[\label{SF:3e}]{%
    \includegraphics[width=0.3\textwidth]{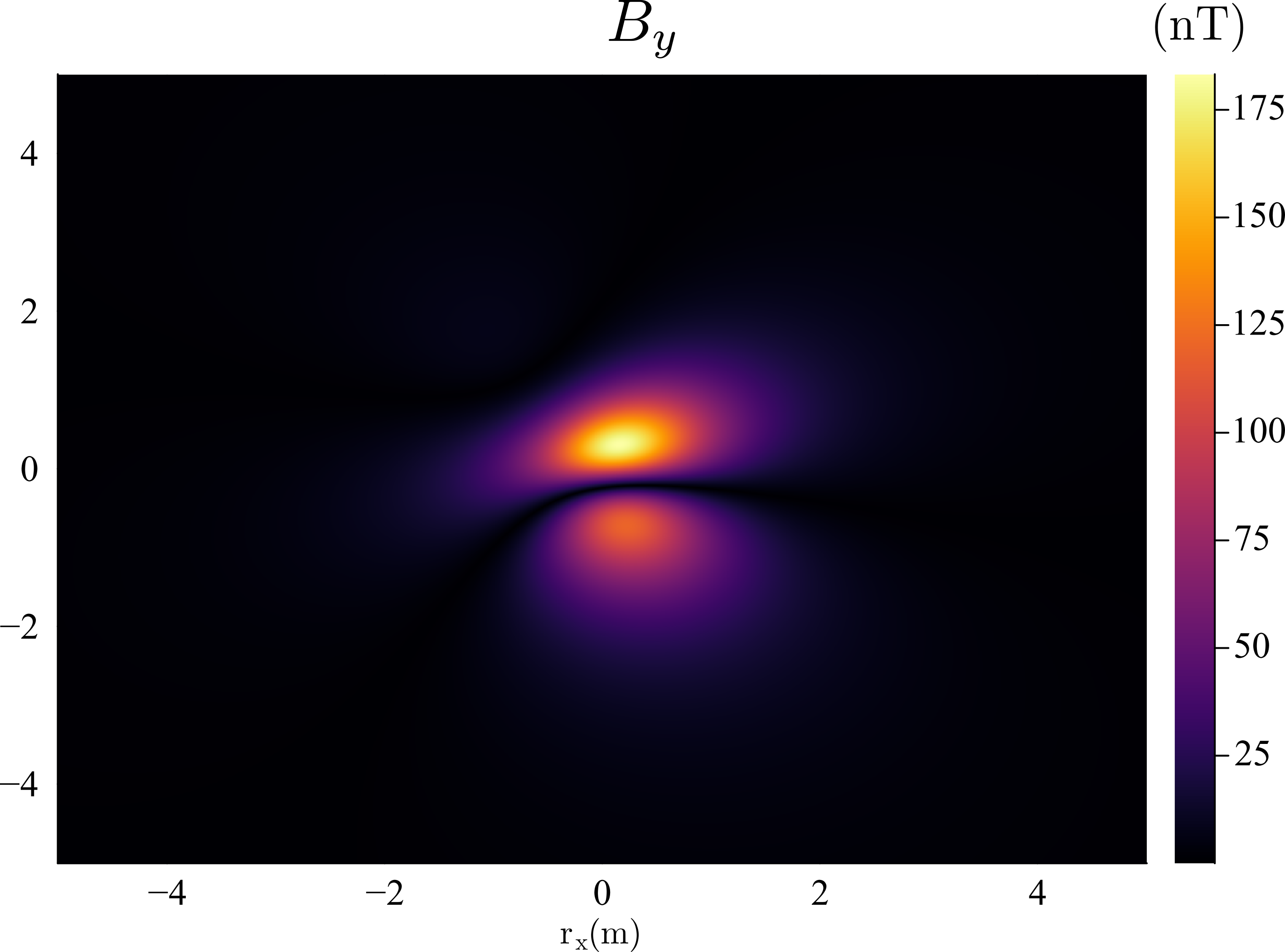}%
    }
    \subfloat[\label{SF:3i}]{%
    \includegraphics[width=0.3\textwidth]{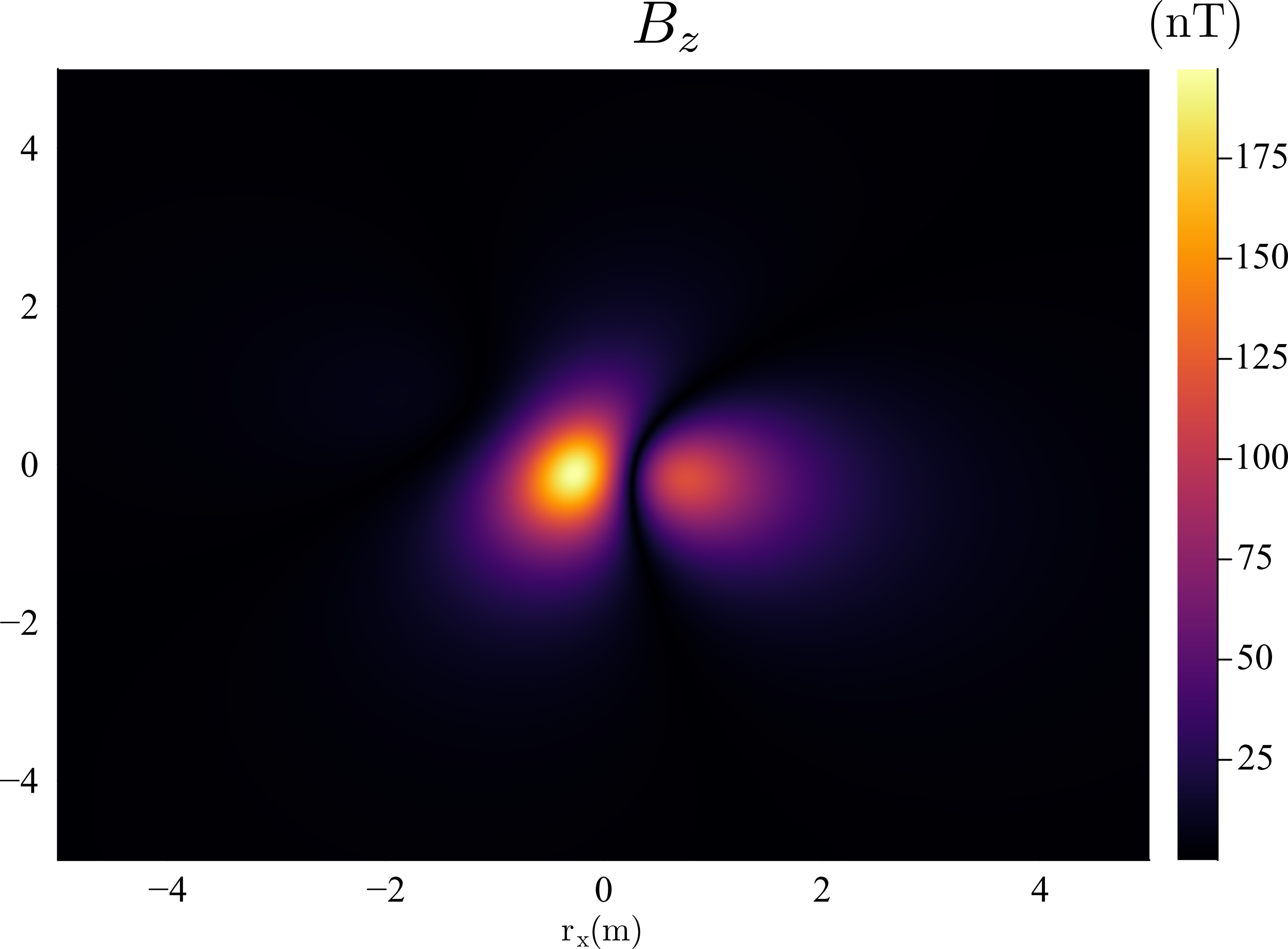}%
    }
    \caption{Magnetic field emitted by a ferromagnetic sphere of radius $R = 16$cm in the far-field regime. The colour plot shows the intensity of the anomaly field (nT) over a square of side $2L = 10$m at a height $h = 1$m. The geomagnetic field that induces the magnetization of the sphere has an intensity $B = 45.85 \ \mathrm{\mu T}$, a colatitude $\theta = 105.63^\circ$ and a longitude $\phi = 320.72^\circ$.}
    \label{SF: MagneticAnomaly}
\end{figure*}

\section{Modeling of magnetic fields from neuron action potentials}\label{SM: NeuronAP}

Neuron action potentials (APs) consist in the depolarization and repolarization of a neuron under a stimulus, which is due to the movement of ions in and out the cell. This increase in the intracellular voltage $\Phi$ of the neuron results in an electric impulse which is transfered to another neuron through an axon, which we model as a wire, such that the current density through the axon is given by Ohm's law:
\begin{equation}
    \mathbf{J} = -\sigma \nabla \Phi(z, \rho, t),
\end{equation}
where $\sigma$ is the conductivity of the axon and $\rho$ is the radial coordinate in cilindrical coordinates. Choosing the direction of propagation of the impulse to be aligned with the $z$ direction, we find the current intensity to be 
\begin{equation}
    I = -\pi\sigma r_a^2 \frac{\partial \Phi(z, \rho, t)}{\partial z},
\end{equation}
with $r_a$ the radius of the axon. Then, assuming a constant conduction velocity, 
\begin{equation}
    \frac{\partial \Phi(z, \rho, t)}{\partial t} = -v_c \frac{\partial \Phi(z, \rho, t)}{\partial z}.
\end{equation}
Finally, using Biot-Savart law, we find the magnetic field associated to the electrical impulse,
\begin{equation}
    B_x(t) = \frac{\mu_0 I}{2\pi \rho} = \frac{\mu_0 \sigma r_a^2}{2 v_c \rho} \frac{\partial \Phi(z, \rho, t)}{\partial t}, 
\end{equation}
where $s \equiv \frac{\mu_0 \sigma r_a^2 }{2 v_c \rho}$ is a proportionality constant. Following \cite{Barry_2016}, we take $s = 7.6 \mathrm{pT/(V/s)}$. Note that we consider the direction of the field at the position of the sensor to be perpendicular to the propagation of the electrical impulse, which we take to be aligned with the $x$ axis. Then, $\rho$ is the distance to the sensor. In this work, we model the intracellular voltage to replicate the experimental results of \cite{Barry_2016}. 

\begin{figure*}[ht]
    \subfloat[\label{SF:4a}]{%
    \includegraphics[width=0.33\textwidth]{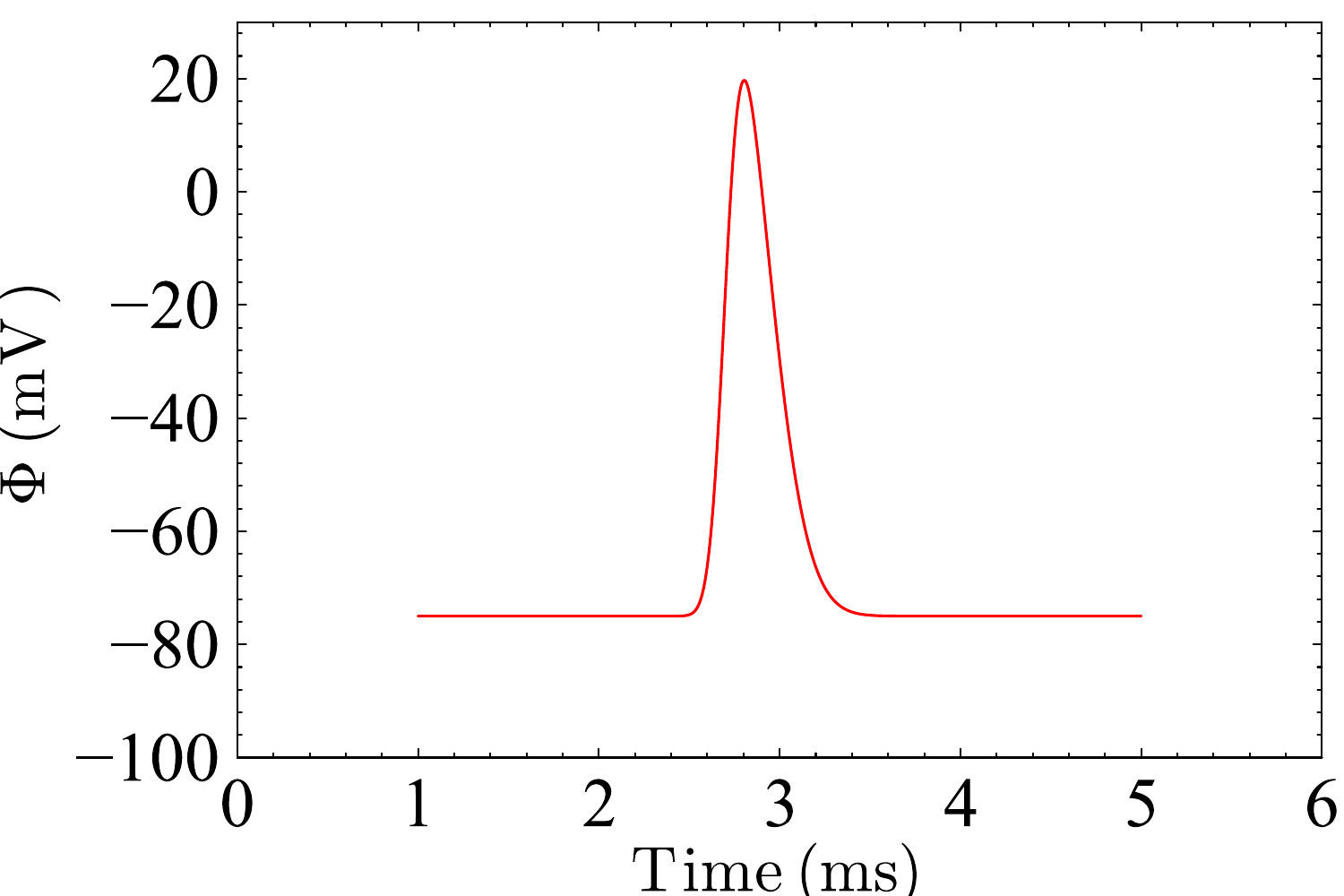}%
    }
    \subfloat[\label{SF:4b}]{%
    \includegraphics[width=0.33\textwidth]{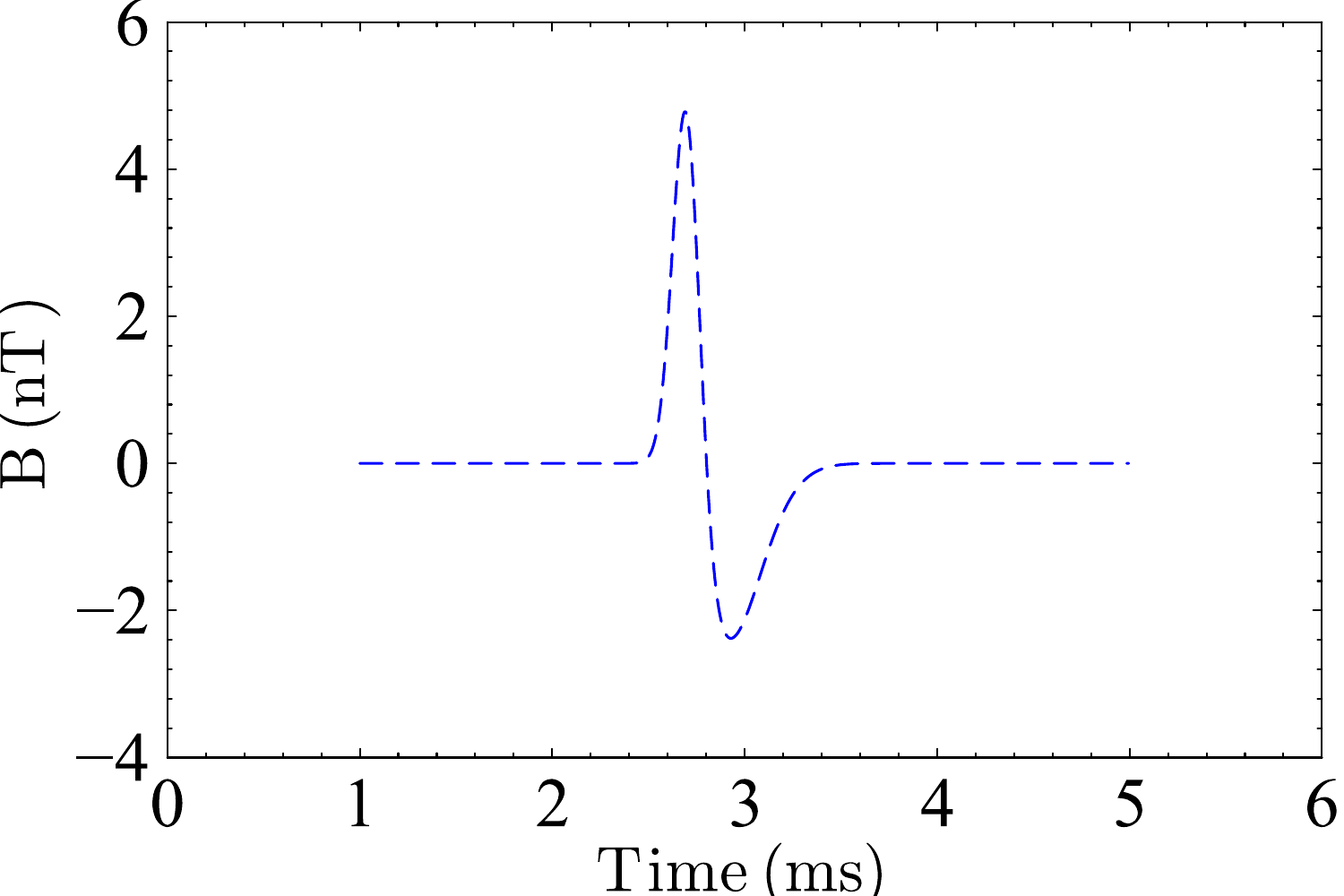}%
    }
    \caption{Gradiometry for magnetophisiology: neuron action potential. (a) Intracellular voltage $\Phi$ that originates the electric impulse. (b) The magnetic field calculated from the intracellular voltage.}\label{SF: NeuronAP}
\end{figure*}



\end{document}